\documentclass{article}
\usepackage[T1,T2A]{fontenc}
\usepackage[utf8]{inputenc}
\usepackage[russian,english]{babel}
\usepackage{amsthm,amssymb,amsmath}
\usepackage{authblk}
\usepackage{graphicx}

\title{Path-integral approach to mutual information calculation for nonlinear communication channel with small second dispersion at large signal-to-noise power ratio}
\author[1,2]{A. V. Reznichenko}
\author[1]{V. O. Guba}
\affil[1]{Theory department, Budker Institute of Nuclear Physics of Siberian Branch Russian
Academy of Sciences, Novosibirsk 630090, Russia}
\affil[2]{Novosibirsk State University, Novosibirsk 630090, Russia}
\date{E-mail: A.V.Reznichenko@inp.nsk.su}

\begin{document}

\maketitle

\begin{abstract}
    We consider the information fiber optical channel modeled by the nonlinear Schrodinger equation with additive Gaussian noise. Using path-integral approach and perturbation theory for the small dimensionless parameter of the second dispersion, we calculate the conditional probability density functional in the leading and next-to-leading order in the dimensionless second dispersion parameter associated with the input signal bandwidth. Taking into account specific filtering of the output signal by the output signal receiver, we calculate the mutual information in the leading and next-to-leading order in the dispersion parameter and in the leading order in the parameter signal-to-noise ratio ($\mathrm{SNR}$). Further, we find the explicit expression for the mutual information in case of the modified Gaussian input signal distribution taking into account the limited frequency bandwidth of the input signal. 
\end{abstract}

\section{Introduction}
Analytical description of information transmission through a nonlinear communication channel remains an open problem of information theory, as it significantly depends on the physical model of the channel, receiver and the decoding procedure. Addressing this problem, one usually aims to calculate the channel capacity, which determines the maximum amount of information per symbol that can be transmitted through a channel. The concept of channel capacity was introduced by Shannon in the paper \cite{shannon1948mathematical}. In the same work he calculated the channel capacity of a linear channel with a Gaussian noise: 
\begin{align}\label{shannonlimit}
   C \propto \log_{2}(1+R_{sn}),
\end{align}
where $R_{sn}$ = $P$/$N$ is the signal-to-noise ratio (SNR), $P$ is the input signal power and $N$ is the noise power. Relation \eqref{shannonlimit} suggests that to increase the channel capacity one should increase the input signal power. This formula is often used for estimations in cases of nonlinear channels or channels with non-Gaussian noise. However, nonlinear effects become significant in fiber-optic communication channels if the signal power grows large, which makes consideration of nonlinear communication channels necessary. 

A model that is widely used to describe signal propagation through a noisy nonlinear optical fiber channel is the nonlinear Schrodinger equation (NLSE) with the additive white Gaussian noise (AWGN) \cite{haus1991quantum}: 
\begin{align}\label{NLSE}
    \partial_{z}\psi + i\beta\partial^2_{t}\psi - i\gamma|\psi|^2\psi = \eta(z,t).
\end{align}
The term $i\gamma |\psi|^{2}\psi$, where $\gamma$ is called the Kerr nonlinearity, in the equation \eqref{NLSE}, is introduced to describe nonlinear effects (self-phase modulation, cross-phase modulation, etc.) due to the Kerr effect, i.e. the effect of changing of the refractive index of the fiber in proportion to the square of the strength of the electric field applied. The term with the second time derivative is responsible for the dispersion, and $\beta$ is called the second dispersion. We assume that the noise function $\eta(z,t)$ in the right-hand side of eq. \eqref{NLSE} has zero mean and finite frequency bandwidth $W'$ of the correlation function: 
\begin{align}
    &\langle\eta(z,t)\rangle_{\eta}=0,\\
    &\langle\eta(z,t)\overline{\eta}(z',t')\rangle_{\eta}=\frac{Q}{\pi (t-t')}\textrm{sin}\left(\frac{W'(t-t')}{2}\right)\delta(z-z'), \label{noisecorrelatortime}
\end{align}
where $Q$ is the noise power per unit length and per unit frequency, and $W'$ is the bandwidth of the noise. The right-hand side of eq. \eqref{noisecorrelatortime} implies that the noise only exists within the frequency interval $[-W'/2,W'/2]$. The brackets $\langle ... \rangle_{\eta}$ mean the averaging over the noise realizations in the channel. The bandwidth $W'$ is introduced to regularize the problem: when $W' \rightarrow +\infty$ one has:
\begin{eqnarray}
    &\langle\eta(z,t)\overline{\eta}(z',t')\rangle_{\eta}\rightarrow Q \delta(t-t')\delta(z-z').
\end{eqnarray}
In this simple model described by the eq. \eqref{NLSE} the noise comes from the equally spaced amplifiers that compensate for the signal attenuation. 

An explicit expression for the capacity of a nonlinear noisy channel with an arbitrary dispersion is not yet known. Considering simplified cases allows one to achieve some insight into the transmission of information through a nonlinear channel. One of such cases is the channel with zero dispersion. When the dispersion is zero, the signals at different time moments propagate independently, which effectively allows one to consider time-independent problem. Such treatment is known as the per-sample model. However, the per-sample model has several limitations. It does not describe spectral broadening of the propagating signal and limited bandwidth of the receiver. Moreover, in the work \cite{kramer2018autocorrelation} it was shown that in the per-sample model with the finite frequency bandwidth of the noise the capacity can reach arbitrarily large values for any input power $P$. Namely, one achieves large rates of information transmission by sending signals at frequencies that belong to the noise-free part of the spectrum. The work \cite{kramer2018autocorrelation} also develops upper bounds for the channel capacity of dispersion-free fiber.

The zero-dispersion case was also considered in papers \cite{turitsyn2003information, terekhov2017log, panarin2017next, reznichenko2019log, reznichenko2022optimal}. In these works the expression for the conditional probability distribution functional (PDF) of the output signal was obtained in the form of functional integral due to the Martin-Siggia-Rose formalism \cite{zinn2021quantum, martin1973statistical}. The conditional PDF, which we denote in our work as $P[Y|X]$, is the probability density to detect the output signal $Y$ if the input signal is $X$. The path-integral approach allows for a perturbative study of the conditional PDF and other information characteristics. The channel capacity of the per-sample model was shown to grow with the average input signal power $P$ as $\textrm{log}\, \textrm{log} \,P$ in the so-called intermediate power range $N \ll P \ll (N\gamma^{2}L^{2})^{-1}$, where $L$ is the channel length and $\gamma$ is the Kerr nonlinearity.  

In order to grasp effects of the dispersion one may consider channels with small nonlinearity and apply the perturbative approach. This approach was used in \cite{shtaif2022challenges} to assess the influence of the nonlinear interference noise on the capacity. The authors obtained upper and lower bounds for the capacity and demonstrated that the difference between the bounds is independent of any parameter of the system in the first order of the perturbation analysis.

In \cite{reznichenko2022optimal} the perturbative study of the nonlinear channel with arbitrary dispersion and large SNR was carried out using the path-integral approach and by the direct calculation of the output signal correlators.

In this paper we address the problem of dispersive communication channels described by the equation \eqref{NLSE} developing perturbative description of the channel with arbitrary Kerr nonlinearity and small second dispersion $\beta$. Saying that $\beta$ is small, we imply in our work smallness of the dimensionless parameter associated with the second dispersion and with the bandwidth of the input signal $X$.

This paper is organized as follows. In the Second section we use the path-integral representation to perturbatively calculate the conditional PDF in the first order in $1/R_{sn}$ and in the leading and the next-to-leading order in the second dispersion. The expression that we obtain for the conditional PDF is valid in the intermediate power range $N \ll P \ll (N\gamma^{2}L^{2})^{-1}$. In the Third section we take into account the limited bandwidth of the receiver. It is achieved by integration of the conditional PDF $P[Y|X]$ over the high-frequency Fourier components of the output signal $Y$ which are not distinguished by the receiver. Then we use the normalization condition to determine the normalization factor of the conditional PDF for the observed components of $Y$. As it was demonstrated in the paper \cite{terekhov2016calculation}, the normalization factor plays a significant role in the calculation of the mutual information. Namely, in the leading order in $1/R_{sn}$ the mutual information is proportional to the logarithm of the normalization factor averaged over realizations of the input signal. The Fourth section contains evaluation of the information characteristics: output signal PDF, conditional entropy, entropy of the output signal and mutual information. The expression for the mutual information is used to evaluate the transmission rate for a specific distribution of the input signal in the Fifth section. Namely, we assumed Gaussian distribution with the limited bandwidth. Such form of the input signal distribution allowed us to use results from the theory of the quantum harmonic oscillator evolving in the imaginary time. The appendices A and B contain details of the calculation of the conditional PDF $P[Y|X]$. The appendix C complements the Fifth section with the details of the calculation of the transmission rate.

\section{Model of the channel}
\subsection{Propagation of the signal}
In our work propagation of the signal $\psi(z,t)$ is governed by the nonlinear Schrodinger equation with additive Gaussian noise \cite{haus1991quantum, agrawal2000nonlinear, menyuk1999application, kodama1985optical}: 
\begin{align}\label{nlse2}
    \partial_{z}\psi + i\beta\partial^2_{t}\psi - i\gamma|\psi|^2\psi = \eta(z,t), 
\end{align}
with the input condition $\psi(z=0,t) = X(t)$. The function $\psi(z,t)$ is related to the modulation of the electric field in the fiber: $\vec{E}(z,t)=\vec{e}_{0}\textrm{Re}\{\psi(z,t+\frac{z}{v_{g}})e^{ik_{0}z-i\omega_{0}t}\}$, where $\omega_{0}$ is the frequency of the carrier wave, $k_{0} = k(\omega_{0})$ is the corresponding wave vector, $v_{g}=\frac{d\omega}{dk}(k_{0})$ is the group velocity and $\vec{e}_{0}$ is the polarization vector. The coefficients $\beta$ and $\gamma$ are the second dispersion and the Kerr nonlinearity, respectively. The equation \eqref{nlse2} does not contain terms responsible for the attenuation of the signal, because in our model we have distributed amplifiers which compensate the attenuation and also cause additive noise $\eta(z,t)$ to appear \cite{haus1991quantum}. In our model of the noise the random function $\eta(z,t)$ has statistical properties of the Gaussian noise with finite frequency bandwidth $W'$:
\begin{align}\label{correlators1}
    &\langle\eta(z,t)\rangle_{\eta}=0,\nonumber\\
    &\langle\eta(z,t)\overline{\eta}(z',t')\rangle_{\eta}=\frac{Q}{\pi (t-t')}\textrm{sin}\left(\frac{W'(t-t')}{2}\right)\delta(z-z'),
\end{align}
where $\langle ... \rangle_{\eta}$ is the averaging over the realizations of the function $\eta(z,t)$, $Q$ is the power of the noise per unit length and per unit frequency, and the bar stands for the complex conjugation. Meaning of the correlators \eqref{correlators1} is transparent in the frequency domain. Namely, for the Fourier transform of the noise function:
\begin{eqnarray}
    \eta(z,\omega) = \int_{-\infty}^{+\infty} dt \textrm{e}^{i\omega t}\eta(z,t)
\end{eqnarray}
the correlation function reads: 
\begin{eqnarray}\label{correlatorfreq}
    \langle\eta(z,\omega)\bar{\eta}(z',\omega') \rangle_{\eta} = 2\pi Q\delta(\omega - \omega') \theta\left(\frac{W'}{2}-|\omega|\right)\delta(z-z'),
\end{eqnarray}
where $\theta(x)$ is the Heaviside theta-function. From the eq. \eqref{correlatorfreq} one sees that the noise is not zero within the frequency interval $\left[-W'/2,W'/2\right]$, and the limit $W' \rightarrow +\infty$ corresponds to the case of white noise. In our work the bandwidth $W'$ plays the role of the ultraviolet cutoff. We are not considering effects of the signals which propagate at such frequencies that $|\omega|>W'/2$.

\subsection{Receiver model}
After the process of propagation the output signal $Y(t) = \psi(z=L,t)$ is detected by a receiver, where $L$ is the distance of propagation. In our work we wish to take into account finite bandwidth $W_{d}$ of the receiver. There may be, in general, different ways of including the receiver bandwidth $W_{d}$ into the model. Finite bandwidth of the receiver implies that high-frequency Fourier harmonics of the output signal $Y(\omega)$ are not distinguished. A straightforward way to get rid of the unobservable degrees of freedom is to integrate the conditional PDF over these high-frequency modes $Y(\omega)$, $W_{d}/2<|\omega|<W'/2$:
\begin{eqnarray}\label{P_domega}
    P_{d}[Y(\omega)|X] = \int_{\{Y(\omega) : |\omega| > \frac{W_{d}}{2}\}} DY P[Y(\omega)|X].
\end{eqnarray}
The new functional $P_{d}[Y(\omega)|X]$ depends only on the observable degrees of freedom and can be used to calculate information characteristics. 

However, one may think of a different detection procedure. For instance, we can choose the receiver  which does not distinguish the high-frequency Fourier harmonics of a different function $\tilde{Y}$, where $\tilde{Y}$ is functionally related to the output signal $Y$. The function $\tilde{Y}$ can be chosen to significantly simplify the analytical consideration, but we get a different functional integrating out the unobservable harmonics of $\tilde{Y}$:
\begin{eqnarray}\label{P_domegatilde}
    \tilde{P}_{d}[Y(\omega)|X] = \int_{\{\tilde{Y}(\omega) : |\omega| > \frac{W_{d}}{2}\}} D\tilde{Y} P[Y(\omega)|X].
\end{eqnarray}

 For the functional $\tilde{P}_{d}$ to be a good approximation of the functional \eqref{P_domega}, it is reasonable to choose $\tilde{Y}$ so that $\tilde{Y}(\omega) \approx Y(\omega)$ at large $|\omega|$. In other words, introduction of $\tilde{P}_{d}$ can be treated both as the model of a new receiver, or as the analytical trick to perform the approximate integration in the eq. \eqref{P_domega}.
\subsection{Input signal}
We consider input signals $X(t)$ with fixed average power $P$ and fixed average bandwidth $W_{X}$, defined as: 
\begin{eqnarray}
    &&P = \int \frac{dt}{T} \langle |X(t)|^{2} \rangle_{P_{X}}, \\
    &&W_{X}^{2} = \int \frac{dt}{T} \langle |\dot{X}(t)|^{2} \rangle_{P_{X}}\bigg/\int \frac{dt}{T} \langle |X(t)|^{2} \rangle_{P_{X}},
\end{eqnarray}
where the brackets $\langle ... \rangle_{P_{X}} = \int DX (...)P_{X}[X]$ are the average with respect to the input signal distribution $P_{X}$. The normalization condition for the input signal distribution reads:
\begin{eqnarray}
    \int DX P_{X}[X] = 1, 
\end{eqnarray}
where the integration measure is $DX(t) = \prod_{i} d\textrm{Re} X(t_{i})d\textrm{Im} X(t_{i})$. We consider such input signals $X(t)$ that vanish outside of the time interval $\left[-T/2, T/2\right]$. We also assume that the input signal is a slowly varying function of time in the sense that $\beta L W_{X}^{2} \ll 1$. The parameters $\beta L W_{X}^{2}$ and $\beta L W_{d}^{2}$ play the roles of the small dimensionless parameters of the perturbative expansion in our work. 

Having introduced all the bandwidths that we work with, we can specify the hierarchy. We assume the following relation on the receiver bandwidth:
\begin{eqnarray}
    \beta L W_{d}^{2} \ll 1,
\end{eqnarray}
and the following hierarchy: 
\begin{eqnarray}\label{hierarchy}
    W_{X} \lesssim W_{d} \ll W'.
\end{eqnarray}

\section{Conditional probability density functional}
\subsection{The path-integral representation of the conditional PDF}
The path-integral approach to the calculation of the conditional PDF $P[Y|X]$ is based on the Martin-Siggia-Rose formalism \cite{terekhov2014conditional}, which allows us to write the functional $P[Y|X]$ in the following form: 
\begin{align}\label{P[Y|X]second}
    P[Y|X] = \int_{\psi(z=0,t)=X(t)}^{\psi(z=L,t)=Y(t)} D\psi \exp\{-S[\psi]/Q\},
\end{align}
where the effective action $S[\psi]$ is associated with the noise statistics and is the integral of the squared left-hand side of the equation \eqref{NLSE}:
\begin{eqnarray}\label{3action}
    S[\psi] = \int_{T} dt \int_{0}^{L} dz \big|\partial_{z}\psi + i\beta \partial_{t}^{2}\psi - i\gamma |\psi|^{2}\psi\big|^{2}.
\end{eqnarray}
When calculating the path integral \eqref{P[Y|X]second}, one should use the retarded discretization scheme in evolution variable $z$ \cite{terekhov2014conditional}: $(\partial_{z}\psi)(z_{n},t) = (\psi(z_{n},t)-\psi(z_{n-1},t))/\delta_{z}$, where $z_{n} = n\delta_{z}$, $\delta_{z}=L/N$ ($z_{0}=0,\,z_{N}=L$). This choice of discretization scheme is motivated by the causal structure of the propagation process \cite{hochberg1999effective, terekhov2014conditional}. The discretization step of time grid is $\delta_{t} = T/M$, and points of the time grid are defined as $t_m = m\delta_t - T/2$ ($t_0 = -T/2$, $t_M=T/2$). The discretization step is larger than $1/W'$. The integration measure $D\psi$ from the eq. \eqref{P[Y|X]second} is defined as:
\begin{eqnarray}
    D\psi = \left(\frac{\delta_t}{\delta_z \pi Q}\right)^{2M} \prod_{i=1}^{N-1} \prod_{j=0}^{M} \left\{\frac{\delta_t}{\delta_z \pi Q}d\textrm{Re} \psi (z_i,t_j) d\textrm{Im} \psi (z_i,t_j)\right\}.
\end{eqnarray}

For the case of large $R_{sn}$, which we consider in our work, it is convenient to factorize the contribution of the classical solution to the Euler-Lagrange equation for the effective action $S[\psi]$:
\begin{eqnarray}\label{P[Y|X]quas}
    P[Y|X] = \Lambda\exp\{-S[\Psi_{cl}]/Q\},
\end{eqnarray}
where $\Psi_{cl}$ is the solution of the equation $\delta S[\Psi_{cl}] = 0$ with the fixed boundary conditions $\Psi_{cl}(z=0,t) = X(t)$ and $\Psi_{cl}(z=L,t) = Y(t)$, and the ''normalization factor'' $\Lambda$ is equal to:
\begin{eqnarray}\label{Lambdaexplicit}
    \Lambda = \int_{\varphi(z = 0, t) = 0}^{\varphi(z = L, t) = 0} D\varphi \exp\left[-\frac{1}{Q}\left\{S[\Psi_{cl} + \varphi] - S[\Psi_{cl}]\right\}\right].
\end{eqnarray}
Large $R_{sn}$ allows us to use the quasiclassical approach, which means that the ''normalization factor'' $\Lambda$ does not depend on properties of output signal Y(t), but depends only on the input signal $X(t)$ in the leading order in $1/R_{sn}$. Therefore, in order to find P[Y|X] we should first calculate the ''classical'' solution $\Psi_{cl}$ and then calculate the normalization factor $\Lambda$ using the normalization condition:
\begin{eqnarray}
    \int DY P[Y|X] = 1.
\end{eqnarray}
As we mentioned before, the normalization factor $\Lambda$ plays a significant role in the calculation of the information characteristics. 
The variational problem $\delta S[\psi] = 0$ which gives the ''classical'' solution $\Psi_{cl}$ can be written explicitly as the following boundary problem: 
\begin{eqnarray}\label{ELproblem}
    \begin{cases}
        &(\partial_{z}+i\beta\partial_{t}^{2}-2i\gamma|\Psi_{cl}|^{2})\mathcal{L}[\Psi_{cl}] + i\gamma\Psi_{cl}^{2}\overline{\mathcal{L}[\Psi_{cl}]} = 0,\\
        &\mathcal{L}[\Psi_{cl}] = (\partial_{z}+i\beta\partial_{t}^{2}-i\gamma|\Psi_{cl}|^{2})\Psi_{cl}, \\
        &\Psi_{cl}(z=0,t) = X(t), \\
        &\Psi_{cl}(z=L,t) = Y(t), \\
    \end{cases}
\end{eqnarray}
where $\overline{\mathcal{L}[\Psi_{cl}]}$ is the complex conjugate of the function $\mathcal{L}[\Psi_{cl}]$. In the large $R_{sn}$ case we can linearize the problem \eqref{ELproblem}. Namely, we present the solution to the problem \eqref{ELproblem} in the form of a small deviation from the solution $\Phi(z,t)$ to the NLSE \eqref{NLSE} without noise: 
\begin{eqnarray}\label{psiclform}
    \Psi_{cl} = \Phi(z,t) + \kappa(z,t)\exp\left\{i\mu\frac{z}{L} + i\phi\right\},
\end{eqnarray}
where $\mu = \gamma L \rho^{2}$, $X = \rho \textrm{e}^{i\phi}$, and the function $\Phi(z,t)$ is the solution to the following problem:
\begin{eqnarray}\label{phiproblem}
    \begin{cases}
        &\partial_{z}\Phi + i\beta\partial^2_{t}\Phi - i\gamma|\Phi|^2\Phi = 0,\\
        &\Phi(z = 0, t) = X(t).
    \end{cases}
\end{eqnarray}
To understand why we can treat the function $\kappa(z,t)$ as a small deviation, one should consider the behaviour of the functional $S[\psi]$ around $\Phi(z,t)$. The action $S[\psi]$ achieves absolute minimum ($S[\Phi]=0$) on the function $\Phi(z,t)$, so the expansion of the functional $S[\Phi + \kappa \exp \left\{i\mu\frac{z}{L} + i\phi\right\} ]$ starts from quadratic terms for small $\kappa$.
Therefore, the conditional PDF $P[Y|X]\propto \exp[-S[\Psi_{cl}]/Q]$ decreases exponentially for such configurations of the output signal $Y(t)$ that make $|\kappa(z,t)|$ much larger than $\sqrt{QLW_{X}}$. The most significant contribution comes from $\kappa(z,t)\propto \sqrt{QLW_{X}}$. To find $P[Y|X]$ in the leading order in $1/R_{sn}$ it will be sufficient to solve the linearized version of the problem \eqref{ELproblem}.

Let us start from calculation of $\Phi(z,t)$ in the form of perturbative expansion in the small second dispersion $\beta$:
\begin{eqnarray}\label{Phidecomposed}
    \Phi = \Phi_{0} + \Phi_{1},
\end{eqnarray}
where $\Phi_{k} \propto \beta^{k}$. Substituting this expansion in eq. \eqref{phiproblem}, we get problems that define terms $\Phi_{0,1}$. For the leading order term we get the following per-sample problem:
\begin{eqnarray}\label{phiproblemzero}
    \begin{cases}
        &\partial_{z}\Phi_{0} - i\gamma|\Phi_{0}|^{2} \Phi_{0} = 0,\\
        &\Phi_{0}(z = 0, t) = X(t),
    \end{cases}
\end{eqnarray}
and then its solution reads: 
\begin{eqnarray}
\Phi_{0}(z,t) = X(t)e^{i\gamma z |X(t)|^{2}} = X(t)e^{i\mu(t)\frac{z}{L}},
\end{eqnarray}
For the next-to-leading order $\Phi_{1}$ we get the following problem: 
\begin{eqnarray}\label{phiproblemone}
    \begin{cases}
        &\partial_{z}\Phi_{1} -i\gamma\Phi^2_{0}\overline{\Phi}_{1} - 2i\gamma|\Phi_{0}|^2\Phi_{1} = -i\beta\partial^2_{t}\Phi_{0},\\
        &\Phi_{1}(z = 0, t) = 0.
    \end{cases}
\end{eqnarray}
This problem has the following solution: 
\begin{eqnarray}
    &\Phi_{1}(z, t)=\beta \exp\left\{i\left(\mu\frac{z}{L}+\phi\right)\right\}\Big\{z[2\dot{\phi}\dot{\rho}+\rho\ddot{\phi}]+z^2\gamma\rho^2[\ddot{\rho}+3\frac{\dot{\rho}^2}{\rho}]+\nonumber\\
    &i\Big(z^3(\gamma\rho^2)^2\frac{2}{3}[\ddot{\rho}+5\frac{\dot{\rho}^2}{\rho}]+z^2\gamma\rho^2[4\dot{\phi}\dot{\rho}+\rho\ddot{\phi}]+z[\rho\dot{\phi}^2-\ddot{\rho}]\Big)\Big\}.
\end{eqnarray}
In order to find the ''classical'' solution $\Psi_{cl}$ and the action $S[\Psi_{cl}]$ we also use the perturbative approach, expanding the function $\kappa(z,t)$ in terms of the small parameter of the second dispersion:
\begin{eqnarray}
    \kappa = \kappa_{0}+\kappa_{1}.
\end{eqnarray}
The linearized problem \eqref{ELproblem} in the leading order in $\beta$ reads: 
\begin{eqnarray}\label{kappa0problem}
    \begin{cases}
        & \partial_{z}^{2}\kappa_{0} - 2i\gamma \rho^{2}\kappa_{0} - 4(\gamma \rho^{2})^{2}\textrm{Re}[\kappa_{0}] = 0, \\
        &\kappa_{0}(z=0,t) = 0, \\
        &\kappa_{0}(z=L,t) = \left[Y(t)-\Phi(z=L,t)\right]\textrm{e}^{-i\phi-i\mu} 
    \end{cases}
\end{eqnarray}
Following \cite{terekhov2017log}, we express the solution to the problem \eqref{kappa0problem} defining new real-valued functions $x(t)$ and $y(t)$ as $x(t) + iy(t) = \left[Y(t)-\Phi(L,t)\right]\textrm{e}^{-i\phi-i\mu}$: 
\begin{eqnarray}\label{kappa0eqs}
    &&\textrm{Re}[\kappa_{0}]=\left(\mu\frac{\mu x - y}{1+\mu^{2}/3}\frac{z}{L}+\frac{(1-2\mu^{2}/3)x+\mu y}{1+\mu^{2}/3}\right)\frac{z}{L},\nonumber\\
    &&\textrm{Im}[\kappa_{0}]=\left(\frac{\mu x - y}{1+\mu^{2}/3}\left[\frac{2\mu^{2}z^{2}}{3L^{2}}-1\right]+\mu\frac{(1-2\mu^{2}/3)x + \mu y}{1+\mu^{2}/3}\right)\frac{z}{L}.
\end{eqnarray}
The function $\kappa_{1}$ is the solution to the problem which is similar to the problem \eqref{kappa0problem}, but with a non-zero right-hand side and with zero boundary conditions: 
\begin{eqnarray}
    \begin{cases}
        & \partial_{z}^{2}\kappa_{1} - 2i\gamma \rho^{2}\kappa_{1} - 4(\gamma \rho^{2})^{2}\textrm{Re}[\kappa_{1}] = F(z,t), \\
        &\kappa_{1}(z=0,t) = 0, \\
        &\kappa_{1}(z=L,t) = 0,
    \end{cases}
\end{eqnarray}
because the condition $\Psi_{cl}(z=L,t) = Y(t)$ is already satisfied by the leading order contribution $\kappa_{0}$. Expression for right-hand side $F(z,t)$ is rather cumbersome, so we present it in the Appendix A: see eq. \eqref{AfunctionF}.

With the expression for $\Psi_{cl}$ we can obtain expressions for the action $S[\Psi_{cl}]=S_{cl}$ in the leading and next to leading orders in parameter $\beta$. The leading order contribution reads: 
\begin{eqnarray}\label{S_0clintegral}
    S_{cl}^{(0)} = \int_{T} dt \frac{(1+4\mu^{2}/3)x^{2} - 2\mu x y + y^{2}}{L(1+\mu^{2}/3)}.
\end{eqnarray}
The first correction to the $S_{cl}^{(0)}$ can be written in the following form:
\begin{eqnarray}\label{S_1clintegral}
    &\!S_{cl}^{(1)} = \beta \int_{T} dt \Big\{a_{1}(t)x^{2} + a_{2}(t)y^{2} + a_{3}(t)xy + a_{4}(t)x\dot{x} + a_{5}(t)y\dot{y} +\nonumber\\
    &\!a_{6}(t)x\dot{y} + a_{7}(t)\dot{x}y + a_{8}(t)x\ddot{x} + a_{9}(t)y\ddot{y} + a_{10}(t)x\ddot{y} + a_{11}(t)\ddot{x}y\Big\},
\end{eqnarray}
where we have introduced functions $a_{i}(t)$ which can be found in the Appendix A.

\subsection{The transformation of the receiver harmonics}

Now that we know the action $S_{cl}$ in the leading and next-to-leading orders in $\beta$, we proceed to take into account the limited bandwidth of the receiver. As we have stated in the section $2.2$, our treatment of the receiver implies integrating the functional $P[Y|X]$ over the high-frequency Fourier harmonics of complex-valued function $\tilde{Y}(t)$. In what follows we choose such a model of the receiver, that the function $\tilde{Y}(t) = y_{1}(t)+iy_{2}(t)$, where $y_{1,2}(t)$ are real-valued, is defined by the following equation: 
\begin{eqnarray}\label{subsYtildeY}
    \frac{y_{1}(t)}{\sqrt{\alpha_{1}}} + i\frac{y_{2}(t)}{\sqrt{\alpha_{2}}} = [Y(t)-\Phi(z=L,t)]e^{i\theta - i\mu - i\phi} = [x(t)+iy(t)]e^{i\theta},
\end{eqnarray}
where $\theta = \frac{1}{2}\arctan{\frac{2}{3\mu}}$, and $\alpha_{1, 2}$ are eigenvalues of the symmetric matrix that represents the quadratic form in the variables $x$ and $y$ from the eq. \eqref{S_0clintegral}: 
\begin{eqnarray}
    S_{cl}^{(0)} = \int dt \frac{1}{L}
    \begin{pmatrix}
        x, y
    \end{pmatrix}
    \begin{pmatrix}
        &\!\!\!\!\frac{1+4\mu^{2}/3}{1+\mu^{2}/3} \quad \frac{-\mu}{1+\mu^{2}/3}\\
        &\!\!\!\!\frac{-\mu}{1+\mu^{2}/3} \quad \frac{1}{1+\mu^{2}/3}
    \end{pmatrix}
    \begin{pmatrix}
        &\!\!\!\!\!\!x\\
        &\!\!\!\!\!\!y
    \end{pmatrix},
\end{eqnarray}
or explicitly: 
\begin{eqnarray}\label{alphaformula}
    &&\alpha_{1} = \frac{3+2\mu^{2}-\mu\sqrt{9+4\mu^{2}}}{3+\mu^{2}},\,\alpha_{2} = \frac{3+2\mu^{2}+\mu\sqrt{9+4\mu^{2}}}{3+\mu^{2}}\label{alphaformula2}.
\end{eqnarray}
In other words, real and imaginary parts of the function $\tilde{Y}(t)$ are related to functions $x(t)$ and $y(t)$ through the following transformation:
\begin{eqnarray}
    \begin{pmatrix}
        x\\
        y
    \end{pmatrix}
    =
    \begin{pmatrix}
        A_{11} & A_{12}\\
        A_{21} & A_{22}
    \end{pmatrix}
    \begin{pmatrix}
    y_{1}\\
    y_{2}
    \end{pmatrix},
\end{eqnarray}
where the coefficients $A_{ij}(t)$ are
\begin{eqnarray}\label{Aij}
    A_{11} = \frac{1}{2} \sqrt{2-\frac{2\mu}{\sqrt{9+4\mu^{2}}}},\quad A_{12} = \frac{1}{2} \sqrt{2+\frac{2\mu}{\sqrt{9+4\mu^{2}}}},\nonumber
\end{eqnarray}
\begin{eqnarray}
    A_{21} = -\sqrt{\frac{1}{2}+\frac{2\mu^{2}}{3}-\frac{\mu\left(15+8\mu^{2}\right)}{6\sqrt{9+4\mu^{2}}}},\quad A_{22} = \sqrt{\frac{1}{2}+\frac{2\mu^{2}}{3}+\frac{\mu\left(15+8\mu^{2}\right)}{6\sqrt{9+4\mu^{2}}}}.
\end{eqnarray}

The way we chose the function $\tilde{Y}(t)$ is dictated by two reasons. First of all, the leading order contribution to the ''classical'' action in terms of the function $\tilde{Y}(t)$ is now diagonal:
\begin{eqnarray}\label{S_cl0integraltilde}
    S_{cl}^{(0)}[\tilde{Y}(t)] = \frac{1}{L}\int_{T} dt \left[y_{1}^{2}(t) + y_{2}^{2}(t)\right].
\end{eqnarray}
The diagonal form of the action \eqref{S_cl0integraltilde} implies that integration of the functional $P[Y|X]$ over Fourier components of the function $\tilde{Y}$ defined by the eq. \eqref{subsYtildeY} is straightforward due to the Parseval's theorem. However, one should keep in mind that the first correction in $\beta$ to the action $S_{cl}$ should be expressed through the functions $y_{1,2}(t)$ as well:
\begin{eqnarray}\label{S_1clintegraltilde}
    &&S_{cl}^{(1)}[\tilde{Y}(t)] = \beta \int_{T} dt \big\{b_{1}y_{1}^{2} + b_{2}y_{2}^{2} + b_{3}y_{1}y_{2} + b_{4}y_{1}\dot{y}_{1} + b_{5}y_{2}\dot{y}_{2} + \nonumber\\
    &&b_{6}y_{1}\dot{y}_{2} + b_{7}\dot{y}_{1}y_{2} + b_{8}y_{1}\ddot{y}_{1} + b_{9}y_{2}\ddot{y}_{2} + b_{10}y_{1}\ddot{y}_{2} + b_{11}\ddot{y}_{1}y_{2}\big\},
\end{eqnarray}
where coefficients $b_{i}(t)$ can be expressed through functions $a_{i}(t)$ (listed in eqs. \eqref{a1}-\eqref{a11}) and $A_{ij}(t)$ (see eq. \eqref{Aij}). The expressions for the functions $b_{i}(t)$ turned out to be rather cumbersome. Fortunately, in order to calculate information characteristics in subsequent sections we only need specific combinations of the functions $b_{i}(t)$. Namely, we only need expressions for $b_{1}(t)+b_{2}(t)$ and $b_{8}(t)+b_{9}(t)$, which turned out to be relatively simple, and we list it in the Appendix B, see eqs. \eqref{Bsumsofbs}. In the appendix B we also show how the functions $b_{i}(t)$ can be expressed in terms of the functions $a_{i}(t)$ and $A_{ij}(t)$.

The second reason for our choice of the function $\tilde{Y}(t)$ is more subtle. We claim that under our assumptions about the input signal, which we stated in the section $2.3$, Fourier components of functions $Y(t)$ and $\tilde{Y}(t)$ become equivalent at high frequencies in the following sense: two conditional PDFs defined by the eq. \eqref{P_domega} and by the eq. \eqref{P_domegatilde} do not differ significantly if the condition \eqref{hierarchy} is satisfied. 

Let us consider the behaviour of Fourier components of functions $Y(t)=Y_{1}(t)+iY_{2}(t)$ and $\tilde{Y}(t)=y_{1}(t)+iy_{2}(t)$. The relation between these functions can be written as:
\begin{eqnarray}\label{Ythroughy}
    Y_{i}(t) = \sum_{j=1,2}B_{ij}(t)y_{j}(t) + \Phi_{i}(t),
\end{eqnarray}
where the matrix $B_{ij}(t)$ is defined by the equation \eqref{subsYtildeY} and $\Phi(z=L,t)=\Phi_{1}(t)+i\Phi_{2}(t)$ with $\Phi_{1}(t)$ and $\Phi_{2}(t)$ being real-valued functions. Explicitly, the matrix $B_{ij}(t)$ reads:
\begin{eqnarray}
    B_{ij} = 
    \begin{pmatrix}
        \alpha_{1}^{-1/2}\cos \left(\theta - \mu - \phi\right) & \alpha_{2}^{-1/2}\sin \left(\theta - \mu - \phi\right)\\
        -\alpha_{1}^{-1/2}\sin \left(\theta - \mu - \phi\right) & \alpha_{2}^{-1/2}\cos \left(\theta - \mu - \phi\right).
    \end{pmatrix}
\end{eqnarray}
An important observation regarding $B_{ij}(t)$ is that this matrix tends to a constant rotation matrix $b_{ij}$ through the angle of $\pi/4$ as time tends to $\pm \infty$, because the input signal $X(t)$ vanishes. Performing the Fourier transform of the equation \eqref{Ythroughy}, we get:
\begin{eqnarray}\label{Yomega}
    Y_{i}(\omega) = \sum_{j=1,2}b_{ij}y_{j}(\omega) + \sum_{j=1,2}\int_{-\infty}^{+\infty} dt F_{ij}(t) y_{j}(t)\textrm{e}^{i\omega t} + \Phi_{i}(\omega),
\end{eqnarray}
where $F_{ij}(t) = B_{ij}(t) - b_{ij}$, $B_{ij}(t) \underset{t \to \pm \infty}{\rightarrow}  b_{ij}$, and the rotation matrix $b_{ij}$ reads: 
\begin{eqnarray}
    \left(b_{ij}\right) = 
    \begin{pmatrix}
        \frac{\sqrt{2}}{2} & \frac{\sqrt{2}}{2}\\
        -\frac{\sqrt{2}}{2} & \frac{\sqrt{2}}{2}
    \end{pmatrix}.
\end{eqnarray}
The Fourier transform is defined for a function of time $f(t)$ as:
\begin{eqnarray}
    f(\omega)=\int_{-\infty}^{+\infty} dt f(t) \textrm{e}^{i\omega t}.
\end{eqnarray}
Now we recall that the input signal vanishes outside of the time interval $[-T/2;T/2]$, so we can replace the integral over time in eq. \eqref{Yomega} with the integral with finite limits. Integrating by parts, we get asymptotic behaviour of $Y_{i}(\omega)$ for high frequencies: 
\begin{eqnarray}
    &&Y_{i}(\omega) \approx \sum_{j=1,2}b_{ij}y_{j}(\omega) + \frac{1}{i\omega}\sum_{j=1,2}\Big(F_{ij}(T/2)y_{j}(T/2)\textrm{e}^{i\omega T/2}+\nonumber\\
    &&F_{ij}(-T/2)y_{j}(-T/2)\textrm{e}^{-i\omega T/2}\Big) + \Phi_{i}(\omega),
\end{eqnarray}
while the asymptotic behaviour of $y_{i}(\omega)$ is:
\begin{eqnarray}
    y_{i}(\omega) \approx \frac{1}{i\omega}\left(y_{i}(T/2)e^{i\omega T/2}+y_{i}(-T/2)e^{-i\omega T/2}\right).
\end{eqnarray}
Due to the factor $F_{ij}(\pm T/2)$ being small we state that at high frequencies $Y_{i}(\omega)$ and $y_{i}(\omega)$ differ only by the rotation $b_{ij}$ and the additive term $\Phi_{i}(\omega)$. The Jacobian determinant of such transformation equals to one, so at high frequencies we have:
\begin{align}
    d\textrm{Re}Y_{1}(\omega)d\textrm{Im}Y_{1}(\omega)d\textrm{Re}Y_{2}(\omega)d\textrm{Im}Y_{2}(\omega) = d\textrm{Re}y_{1}(\omega)d\textrm{Im}y_{1}(\omega)d\textrm{Re}y_{2}(\omega)d\textrm{Im}y_{2}(\omega).
\end{align}

However, the fact that at high frequencies we can change the integration variables from $Y_{1,2}(\omega)$ to $y_{1,2}(\omega)$ does not imply that the detection procedures defined by the eqs. \eqref{P_domega} and \eqref{P_domegatilde} will give the same results. Namely, due to the relation \eqref{Yomega} values of $Y_{1,2}(\omega)$ at high frequencies depend on values of $y_{1,2}(\omega)$ at low frequencies as well:
\begin{eqnarray}\label{Yomegayomega}
    Y_{i}(\omega) = \sum_{j=1,2}\int \frac{d\omega'}{2\pi}B_{ij}(\omega-\omega')y_{j}(\omega') + \Phi_{i}(\omega).
\end{eqnarray}
From the eq. \eqref{Yomegayomega} it follows that in general the result of integration depends on whether we keep fixed the low-frequency values of $Y_{1,2}(\omega)$ or $y_{1,2}(\omega)$. These are related through the matrix $B_{ij}(\omega)$, which inherits its frequency bandwidth from the input signal $X(t)$. In our work we consider the input signal to be a slowly varying function with a small bandwidth, which implies that the value of $B_{ij}(\omega)$ decreases rapidly as $|\omega|$ increases. Therefore, in our case the relation between $Y_{1,2}(\omega)$ and $y_{1,2}(\omega')$ is negligible if $|\omega-\omega'|$ is large. Due to this fact, we expect the conditional PDF defined by the eq. \eqref{P_domegatilde} to be a good approximation of the conditional defined by a more straightforward procedure \eqref{P_domega}.

\subsection{The result of the calculation of $P_{d}[Y|X]$}

The details of how we calculated the integral from the eq. \eqref{P_domegatilde} are presented in the Appendix B. Here we only write out the result: 

\begin{eqnarray}\label{Pdtildet}
    &&\tilde{P}_{d}[Y_{d}|X] = \tilde{\Lambda}_{d}^{(0)} \exp\left\{-\frac{\tilde{\delta}_{t}}{QL}\sum_{j=0}^{M_{d}-1} \left[y_{1}^{2}(t_{j}) + y_{2}^{2}(t_{j})\right]\right\}\Bigg(1 + \frac{\tilde{\Lambda}_{d}^{(1)}}{\tilde{\Lambda}_{d}^{(0)}} -\nonumber\\
    && \frac{\beta\tilde{\delta}_{t}}{Q} \sum_{j=0}^{M_{d}-1} \{b_{1,j}y_{1,j}^{2} + b_{2,j}y_{2,j}^{2} + b_{3,j}y_{1,j}y_{2,j} + b_{4,j}y_{1,j}(\dot{y}_{1})_{j}  + \nonumber\\
    && b_{5,j}y_{2,j}(\dot{y}_{2})_{j} +b_{6,j}y_{1,j}(\dot{y}_{2})_{j} + b_{7,j}(\dot{y}_{1})_{j}y_{2,j} + b_{8,j}y_{1,j}(\ddot{y}_{1})_{j}+ \nonumber\\
    && b_{9,j}y_{2,j}(\ddot{y}_{2})_{j} + b_{10,j}y_{1,j}(\ddot{y}_{2})_{j} + b_{11,j}(\ddot{y}_{1})_{j}y_{2,j}\}\Bigg).
\end{eqnarray}
In the eq. \eqref{Pdtildet} we introduced a discretized time grid with the discretization step $\tilde{\delta}_{t}$ related to the bandwidth of the receiver: $\tilde{\delta}_{t}=2 \pi/W_{d}$. We call this grid coarse, because its discretization step is much larger than the discretization step $2\pi/W'$ related to the bandwidth of the noise $W'$. The functions $y_{1,2}(t)$ now contain only observable Fourier components with frequencies inside of the interval $[-W_{d}/2,W_{d}/2]$, and we reflect it by writing the argument of the PDF as $Y_{d}$. In the normalization factor, which we denote as $\tilde{\Lambda}_{d}$, we separated leading order and next-to-leading order in $\beta$ and wrote it as $\tilde{\Lambda}_{d} = \tilde{\Lambda}^{(0)}_{d}+\tilde{\Lambda}^{(1)}_{d}$. We calculated the normalization factor using the condition: 
\begin{eqnarray}
    \int DY_{d} \tilde{P}_{d}[Y_{d}|X] = 1.
\end{eqnarray}
The normalization factor depends on the input signal and has the following form: 
\begin{eqnarray}\label{tildelambda}
    && \tilde{\Lambda}_{d}^{(0)} = \left(\frac{\tilde{\delta}_{t}}{\pi Q L}\right)^{M_{d}}, \nonumber \\
    &&\frac{\tilde{\Lambda}_{d}^{(1)}}{\tilde{\Lambda}_{d}^{(0)}} = M_{d}\frac{\beta L W_{d}^{2}}{12} \int_{T} \frac{4 \mu^{3}}{15(3+\mu^{2})} \frac{dt}{T} - M_{d} \beta L \int_{T} \Bigg[\frac{4\mu\dot{\mu}\dot{\phi}}{3+\mu^{2}} +\nonumber\\
    && \frac{2(3+2\mu^{2})\ddot{\phi}}{3+\mu^{2}} + \frac{\mu\dot{\mu}^{2}}{15(3+\mu^{2})^{3}(9+4\mu^{2})^{2}}(10206 + 21303\mu^{2} +\nonumber \\
    &&15399\mu^{4}+4644\mu^{6} +496\mu^{8})\Bigg]\frac{dt}{T},
\end{eqnarray}
where the integration over time should be understood as the summation over points of the coarse grid as in the eq. \eqref{Pdtildet}. However, the input signal $X(t)$ does not change significantly over the period of $\tilde{\delta}_{t}$ (because $W_{d} \gg W_{X}$), so we can replace the sum with the continuous limit. 

As we can see in the eqs. \eqref{Pdtildet} and \eqref{tildelambda}, the first correction to the functional $\tilde{P}_{d}[Y_{d}|X]$ is proportional to the small dimensionless parameters of dispersion (which are $\beta L W_{X}^{2}$, $\beta L W_{d}^{2}$ and $\beta L W_{X}W_{d}$) and does not depend on the noise bandwidth $W'$. One may ask whether we can be sure if the contributions of higher orders of the perturbative expansion in $\beta$ will not contain terms which grow infinitely with the noise bandwidth $W'$. To address this question, we estimated the second order correction to the functional $\tilde{P}_{d}[Y_{d}|X]$. We found out that the terms from the second order correction are all bounded.

\section{Mutual information}

In this section we calculate information characteristics using the PDF $\tilde{P}_{d}[\tilde{Y}_{d}|X]$. First of all, we can rewrite the formula \eqref{Pdtildet} in a more compact way:
\begin{eqnarray}
    &&\tilde{P}_{d}[Y_{d}|X] = \left(\tilde{\Lambda}_{d}^{(0)}+\tilde{\Lambda}_{d}^{(1)}\right)\left(1-\frac{S_{coarse}^{(1)}}{Q}\right)\exp\left\{-\frac{S_{coarse}^{(0)}}{Q}\right\}\approx\nonumber \\
    &&\left(\tilde{\Lambda}_{d}^{(0)}+\tilde{\Lambda}_{d}^{(1)}\right)\exp\left\{-\frac{S_{coarse}^{(0)}+S_{coarse}^{(1)}}{Q}\right\},
\end{eqnarray}
where the second equality is true if we neglect all the terms with powers of $\beta$ higher than one. We denoted the action functional on the coarse grid as $S_{coarse}$:
\begin{eqnarray}
    &&S_{coarse}^{(0)} = \frac{\tilde{\delta}_{t}}{L}\sum_{j=0}^{M_{d}-1} \left[y_{1}^{2}(t_{j}) + y_{2}^{2}(t_{j})\right],\\
    &&S_{coarse}^{(1)} = \beta\tilde{\delta}_{t} \sum_{j=0}^{M_{d}-1} \big\{b_{1,j}y_{1,j}^{2} + b_{2,j}y_{2,j}^{2} + b_{3,j}y_{1,j}y_{2,j} + b_{4,j}y_{1,j}(\dot{y}_{1})_{j}  + \nonumber\\
    && b_{5,j}y_{2,j}(\dot{y}_{2})_{j} +b_{6,j}y_{1,j}(\dot{y}_{2})_{j} + b_{7,j}(\dot{y}_{1})_{j}y_{2,j} + b_{8,j}y_{1,j}(\ddot{y}_{1})_{j}+ \nonumber\\
    && b_{9,j}y_{2,j}(\ddot{y}_{2})_{j} + b_{10,j}y_{1,j}(\ddot{y}_{2})_{j} + b_{11,j}(\ddot{y}_{1})_{j}y_{2,j}\big\}.
\end{eqnarray}
We begin the calculation of the information characteristics with the conditional entropy of our channel, defined as follows:
\begin{eqnarray}\label{H[Ytilde|X]}
H_{Y|X} = - \int DX P_{X}[X] \left\{\int D\tilde{Y}_{d} \tilde{P}_{d}[\tilde{Y}_{d}|X] \ln \tilde{P}_{d}[\tilde{Y}_{d}|X] \right\}.
\end{eqnarray}
The conditional entropy is the measure of the noise impact on the propagation of the signal --- it is a negative contribution to the information transmitted through the channel.
The integral over $\tilde{Y}_{d}$ in the eq. \eqref{H[Ytilde|X]} consists of two terms:
\begin{eqnarray}\label{curbrackets}
 &\int D\tilde{Y}_{d} \tilde{P}_{d}[\tilde{Y}_{d}|X] \ln \left[  \left(\tilde{\Lambda}_{d}^{(0)} + \tilde{\Lambda}_{d}^{(1)}\right) \exp \left\{- \frac{S_{coarse}^{(0)} + S_{coarse}^{(1)}}{Q} \right\} \right] = \nonumber\\
 & = \ln [\tilde{\Lambda}_{d}^{(0)} + \tilde{\Lambda}_{d}^{(1)}] - \int D\tilde{Y}_{d} \left( \frac{S_{coarse}^{(0)} + S_{coarse}^{(1)}}{Q}\right) \tilde{P}_{d}[\tilde{Y}_{d}|X],
\end{eqnarray}
where we have used the normalization condition $\int D\tilde{Y}_{d} \tilde{P}_{d}[\tilde{Y}_{d}|X] = 1$ and the fact that the normalization factor does not depend on the output signal $\tilde{Y}_{d}$. The remaining integral in the eq. \eqref{curbrackets} is gaussian and can be evaluated by using the Wick's theorem for gaussian integrals: 
\begin{eqnarray}
    \int D\tilde{Y}_{d} \left( \frac{S_{coarse}^{(0)} + S_{coarse}^{(1)}}{Q}\right) \tilde{P}_{d}[\tilde{Y}_{d}|X] = M_{d}.
\end{eqnarray}
Thus, the conditional entropy is expressed through the logarithm of the normalization constant averaged over $P_{X}$: 
\begin{eqnarray}
    H_{Y|X} = -  \langle\ln [\tilde{\Lambda}_{d}^{(0)} + \tilde{\Lambda}_{d}^{(1)}]\rangle_{P_{X}} + M_{d}.
\end{eqnarray}

In order to calculate the mutual information, which is defined as:
\begin{eqnarray}
    I_{P_{X}} = H_{Y} - H_{Y|X},
\end{eqnarray}
we also need the entropy of the output signal $H_{Y}$, which is expressed through the output signal PDF: 
\begin{eqnarray}
    &P_{out}[\tilde{Y}_{d}] = \int DX P_{X}[X] \tilde{P}_{d}[\tilde{Y}_{d}|X], \\
    &H_{Y} = - \int D\tilde{Y}_{d} P_{out}[\tilde{Y}_{d}] \ln P_{out}[\tilde{Y}_{d}].\label{H[Ytilde]}
\end{eqnarray} 
Since the average noise power is much less than the average input signal power ($P \gg QLW_{X}$), we can use the Laplace method to obtain the following result in the leading order in $1/R_{sn}$ \cite{terekhov2016calculation}:

\begin{eqnarray}\label{Pouttilde}
    P_{out}[\tilde{Y}_{d}] \approx \mathbb{J}_{d}[\Phi^{-1}(Y_{d})] P_{X}[\Phi^{-1}(Y_{d})],
\end{eqnarray}
here $\Phi^{-1}(Y)$ is the nonlinear function which recovers the input condition $\Phi(z = 0, t) = \Phi^{-1}(Y)$ from the corresponding solution of the NLSE without noise taken at the point $z = L$, $Y(t) = \Phi(z = L, t)$. We denoted the Jacobian of the transformation \eqref{subsYtildeY} on the coarse grid by $\mathbb{J}_{d}[X]$:
\begin{eqnarray}
    \mathbb{J}_{d}[X] = \prod_{j=0}^{M_{d} - 1} \sqrt{1+\mu_{j}^{2}/3}.
\end{eqnarray}
Now we return to the eq. \eqref{H[Ytilde]} and make the following substitution: 
\begin{eqnarray}\label{backsubstitution}
    X = \Phi^{-1}(Y).
\end{eqnarray}
The Jacobian of the substitution \eqref{backsubstitution} equals one due to the Liouville's theorem for Hamiltonian systems \cite{novikov1984theory}. In the leading order in $1/R_{sn}$ for the entropy of the output signal we get: 
\begin{eqnarray}
    &&H_Y = H_X - \langle\ln\mathbb{J}_{d}[X]\rangle_{P_{X}} = \nonumber\\
    &&=H_X - M_{d}\left\langle\int_{T}\frac{dt}{T}\ln{\sqrt{1+\mu^{2}/3}}\right\rangle_{P_{X}}.
\end{eqnarray}

We get the mutual information subtracting the conditional entropy \eqref{H[Ytilde|X]} from the entropy of the output signal \eqref{H[Ytilde]}: 
\begin{eqnarray}
    &&I_{P_{X}} = H_X - M_{d}\left\langle\int_{T}\frac{dt}{T}\ln{\sqrt{1+\mu^{2}/3}}\right\rangle_{P_{X}} -\nonumber\\
    &&M_{d} + \langle\ln [\tilde{\Lambda}_{d}^{(0)} + \tilde{\Lambda}_{d}^{(1)}]\rangle_{P_{X}}.
\end{eqnarray}
After substituting the explicit expressions for $\tilde{\Lambda}_{d}^{(0)}$ and $\tilde{\Lambda}_{d}^{(1)}$ from the eq. \eqref{tildelambda} we get the following result: 
\begin{eqnarray}\label{mutualinformation}
    &&I_{P_{X}} = H_X - M_{d}\left\langle\int_{T}\frac{dt}{T}\ln{\sqrt{1+\mu^{2}/3}}\right\rangle_{P_{X}} + M_{d}\Bigg\langle\Bigg\{\ln\left[\frac{\tilde{\delta_{t}}e^{-1}}{\pi Q L}\right] +  \nonumber\\
    &&\frac{\beta L W_{d}^{2}}{12} \int_{T}\frac{dt}{T} \frac{4\mu^{3}}{15(3+\mu^{2})} - \beta L \int_{T}\frac{dt}{T} \Bigg[\frac{4\mu\dot{\mu}\dot{\phi}}{3+\mu^{2}} + \frac{2(3+2\mu^{2})\ddot{\phi}}{3+\mu^{2}} + \nonumber \\
    && \frac{\mu\dot{\mu}^{2}}{15(3+\mu^{2})^{3}(9+4\mu^{2})^{2}}(10206 + 21303\mu^{2} +15399\mu^{4}+\nonumber\\
    &&4644\mu^{6} +496\mu^{8})\Bigg] \Bigg\}\Bigg\rangle_{P_{X}}.
\end{eqnarray}

\section{Mutual information for the Gaussian input signal distribution with the restrained signal bandwidth}
In this section to get understanding of the behaviour of the mutual information from the eq. \eqref{mutualinformation} we wish to evaluate it with respect to the Gaussian distribution of the input signal $P_{X}$, which is the optimal distribution of a linear channel: 
\begin{eqnarray}\label{P0input}
    P_{X}^{(0)}[X] = \Lambda_{X}^{(0)} \exp \left[-\frac{1}{P}\int_{T} \frac{dt}{T} |X|^{2}\right],
\end{eqnarray}
where $P$ is assumed to be fixed:
\begin{eqnarray}
    \int_{T} \frac{dt}{T}\left\langle|X|^{2}\right\rangle_{P_{X}^{(0)}} = P.
\end{eqnarray}
However, we wish to consider input signals of the fixed average bandwidth $W_{X}$, while the exponent in the eq. \eqref{P0input} does not contain any terms which suppress fast realizations of the input signal (i.e. with large $W_{X}$). For this reason we modify the distribution \eqref{P0input} by adding the term proportional to the squared derivative of the input signal:
\begin{eqnarray}\label{PXosc}
    P_{X}[X] = \Lambda_{X} \exp \left[-\frac{1}{P}\int_{T} \frac{dt}{T} \left\{\left(\frac{T}{2\xi} \right)^{2} |\dot{X}|^{2}+|X|^{2}\right\}\right],
\end{eqnarray}
where $\xi$ is the real number chosen so that the average power remains to be equal to $P$:
\begin{eqnarray}\label{powerintegral}
    && \int_{T} \frac{dt}{T} \langle |X|^{2} \rangle_{P_{X}} = P.
\end{eqnarray}
In other words, the condition \eqref{powerintegral} gives us the equation on $\xi$ which reads:
\begin{eqnarray}\label{xiequation}
    3 = 2\xi\, \textrm{cth}\,2\xi.
\end{eqnarray}
The equation \eqref{xiequation} can be solved numerically. The value of $\xi$ is roughly $1.49$. The factor $\Lambda_{X}$ is the normalization factor which ensures that the normalization condition is satisfied: 
\begin{eqnarray}
    \int DX P_{X}[X] = 1.
\end{eqnarray}
The average bandwidth of the input signal, as we defined it in the second section, reads: 
\begin{eqnarray}\label{widthintegral}
    && W_{X}^{2} = \frac{1}{P} \int_{T} \frac{dt}{T} \langle |\dot{X}|^{2} \rangle_{P_{X}}.
\end{eqnarray}
We also recall that the input signal vanishes at the boundaries of the time interval $[-T/2, T/2]$: 
\begin{eqnarray}\label{boundaryconditions}
    X(-T/2) = X(T/2) = 0.
\end{eqnarray}

The exponent in the eq. \eqref{PXosc} is essentially the action of the quantum oscillator with imaginary time, which can be made explicit if we rename the coefficients in the eq. \eqref{PXosc}:
\begin{eqnarray}\label{momeganotation}
    \frac{1}{\xi P} = m \Omega, \, 2\xi=\Omega T.
\end{eqnarray}
Then the input signal distribution reads: 
\begin{eqnarray}\label{PXthroughS}
    P_{X}[X] = \Lambda_{X}\textrm{e}^{-S[X]},
\end{eqnarray}
and for the analogy to be completely evident we also mention that the oscillator is considered in such a system of units that $\hbar = 1$. The action $S[X]$ is:
\begin{eqnarray}\label{SXactionosc}
    S[X] = \int_{T} dt \left(\frac{m |\dot{X}|^{2}}{2}+\frac{m\Omega^{2}|X|^{2}}{2}\right).
\end{eqnarray}
With the action \eqref{SXactionosc} we will be able to use well-known expressions from statistical mechanics for averages with respect to the distribution \eqref{PXthroughS}. 

Calculating averages over $P_{X}$ in the eq. \eqref{mutualinformation} with respect to the distribution \eqref{PXthroughS}, one has to choose a discretization scheme for derivatives. In the Appendix C we use the following discretization for the derivative operator $D_{ij}$:
\begin{eqnarray}\label{generaldisc}
    \dot{g}_{i}=\sum_{j=0}^{M}D_{ij}g_{j} = \dfrac{1}{\delta_{t}}\left(g_{i+1}-g_{i}\right), \, 0 \leq i \leq M-1,
\end{eqnarray}
where $g_{i} = g(t_{i}), \, \dot{g}_{i} = \dot{g}(t_{i})$  and $t_{i}$ is a point of the time grid with the discretization step $\delta_{t}$: $t_{i} = -T/2 + i\delta_{t}$. Evaluation of the average power and average bandwidth with the discretization scheme \eqref{generaldisc} gives: 
\begin{eqnarray}
    && P = \frac{1}{m\Omega}\frac{\Omega T\,\textrm{cth}\,\Omega T - 1}{\Omega T}, \\ \nonumber
    && W_{X}^{2} = \dfrac{2\delta_{t}}{PT}\sum_{\alpha=1}^{M-1}\dfrac{\nu_{\alpha}}{Q_{\alpha}} = \dfrac{2(M-1)}{mPT} - \dfrac{2\Omega^{2}}{mPT}\sum_{\alpha=1}^{M-1}\dfrac{1}{\nu_{\alpha}+\Omega^{2}},
\end{eqnarray}
where we denoted by $\nu_{\alpha}$ eigenvalues of the matrix $(D^{T}D)_{ij}$ (see eq. \eqref{Cnualpha} for the explicit expression for $\nu_{\alpha}$). We remind that the operator $D_{ij}$ acts on the functions with boundary conditions \eqref{boundaryconditions}. The expression for $W_{X}^{2}$ turns out to be divergent if we try to pass to the continuous limit, but the same singularity will appear in the evaluation of the mutual information \eqref{mutualinformation}, where we will express the singular term through the average bandwidth $W_{X}$ and the average power $P$ to obtain finite result. In other words, the equation \eqref{widthintegral} plays the role of the renormalization condition. 

First of all, let us consider averages from the eq. \eqref{mutualinformation} which contain derivatives of the phase $\phi$ of the input signal. If we substitute $X(t) = \rho(t) \textrm{e}^{i\phi(t)}$ into the action \eqref{SXactionosc}, we get: 
\begin{eqnarray}
    S[X] = \int_{T} dt \left(\frac{m |\dot{X}|^{2}}{2}+\frac{m\Omega^{2}|X|^{2}}{2}\right) = \int_{T} dt \Bigg(\frac{m}{2}(\dot{\rho}^{2}+\rho^{2}\dot{\phi}^{2})+\frac{m\Omega^{2} \rho^{2}}{2}\Bigg).
\end{eqnarray}
We see that average of any expression linear in $\phi$ vanishes, because the action is quadratic in $\phi$.

Now we have two terms which make up the first order correction in $\beta$ to the mutual information: 
\begin{eqnarray}\label{firstintegral5}
    &&\int_{T} \frac{dt}{T}\left\langle \frac{4 \mu^{3}}{15(3+\mu^{2})} \right\rangle_{P_{X}}
\end{eqnarray}
and
\begin{eqnarray}\label{fintegral5}
    &&\int_{T} \frac{dt}{T} \langle f(\mu) \dot{\mu}^{2} \rangle_{P_{X}},
\end{eqnarray}
where we denoted
\begin{eqnarray}
    && f(\mu) = \frac{\mu}{15(3+\mu^{2})^{3}(9+4\mu^{2})^{2}}(10206 + 21303\mu^{2}+15399\mu^{4}+\\\nonumber
    &&4644\mu^{6}+496\mu^{8}). \nonumber
\end{eqnarray}
The details of how we evaluated the integrals from the eqs. \eqref{firstintegral5} and \eqref{fintegral5} are presented in the appendix C. Eventually, we obtain following results: 
\begin{eqnarray}\label{averagesresult}
    &&\int_{T} \frac{dt}{T}\langle f(\mu)\dot{\mu}^{2} \rangle_{P_{X}} \approx (\gamma L P )^{2} W_{X}^{2}\int_{0}^{+\infty}dy y f\left(\gamma L P\xi\,  y\right)\\ \nonumber
    &&K_{0}\left(y/\textrm{sh}\,2\xi\right)\textrm{e}^{-y\, \textrm{cth}\,2\xi}, \\ \nonumber
    &&  \int_{T} \frac{dt}{T}\left\langle \frac{4 \mu^{3}}{15(3+\mu^{2})} \right\rangle_{P_{X}} = \frac{2}{15}\xi^{2}(\gamma L P)^{3}\int_{0}^{+\infty}dy \frac{y^{3}K_{0}\left(y/\textrm{sh}\,2\xi\right)}{3+(\gamma L P\xi)^{2}\,y^{2}}\textrm{e}^{-y\,\textrm{cth}\, 2\xi},
\end{eqnarray}
where $K_{0}(x)$ is the modified Bessel function of the second kind. In the first integral in the eq. \eqref{averagesresult} we omitted terms which are proportional to $1/T^{2}$ instead of $W_{X}^{2}$, because we assume that $W_{X}^{2}\gg1/T^{2}$.

Substituting the averages \eqref{averagesresult} into the eq. \eqref{mutualinformation}, we obtain the mutual information with the first-order correction in $\beta$:
\begin{eqnarray}
    &&I_{P_{X}} = H_X - M_{d}\left\langle\int\frac{dt}{T}\ln{\sqrt{1+\mu^{2}/3}}\right\rangle_{P_{X}} + M_{d}\ln\left[\frac{\tilde{\delta_{t}}e^{-1}}{\pi Q L}\right]\nonumber \\
    && + M_d\beta L W_{d}^{2} I_{d}(\tilde{\gamma}) - M_d\beta L W_{X}^{2} I_{X}(\tilde{\gamma}),
\end{eqnarray}
where we denoted the dimensionless nonlinearity parameter as $\tilde{\gamma} \equiv \gamma L P$. The functions $I_{d}$ and $I_{X}$ are defined as:
\begin{eqnarray}\label{Iddefinition}
    I_{d}(\tilde{\gamma}) = \frac{1}{12} \int_{T} \frac{dt}{T}\left\langle \frac{4 \mu^{3}}{15(3+\mu^{2})} \right\rangle_{P_{X}},
\end{eqnarray}
and 
\begin{eqnarray}\label{IXdefinition}
    I_{X}(\tilde{\gamma}) = \frac{1}{W_{X}^{2}} \int_{T} \frac{dt}{T}\langle f(\mu)\dot{\mu}^{2} \rangle_{P_{X}}. 
\end{eqnarray}

Now we wish to investigate the behaviour of the first-order correction to the mutual information as a function of the dimensionless nonlinearity parameter $\tilde{\gamma}$ and the ratio $W_{d}/W_{X}$. For that goal, it is convenient to introduce some more notation, denoting the first-order correction divided by $M_{d}$ as $\Delta I_{P_{X}}$:
\begin{eqnarray}
    \Delta I_{P_{X}} = \beta L W_{d}^{2} I_{d}(\tilde{\gamma}) - \beta L W_{X}^{2} I_{X}(\tilde{\gamma}).
\end{eqnarray}
It is also convenient to introduce a dimensionless function $G(\tilde{\gamma},v)$, which we define by the following relation:
\begin{eqnarray}
    \Delta I_{P_{X}} = \beta L W_{X}^{2}G(\tilde{\gamma}, W_{d}/W_{X}),
\end{eqnarray}
so the function $G(\tilde{\gamma}, v)$ reads: 
\begin{eqnarray}
    G(\tilde{\gamma}, v) = v^{2}I_{d}(\tilde{\gamma}) - I_{X}(\tilde{\gamma}).
\end{eqnarray}

As we shall see shortly, the behaviour of $G(\tilde{\gamma}, v)$ as a function of $\tilde{\gamma}$ significantly depends on the value of $v$. However, we begin with demonstrating functions $I_{d}(\tilde{\gamma})$ and $I_{X}(\tilde{\gamma})$ separately in fig. 1 and fig. 2, respectively. To show the contributions each of these functions gives to the mutual information, we have to choose such values for the parameters $\beta$, $L$, $W_{d}$ and $W_{X}$ that the dimensionless parameters $\beta L W_{d}^{2}$ and $\beta L W_{X}^{2}$ are both small. Let us put, for demonstration, $\beta = 2 \cdot 10^{-23}\,s^{2}/km,\, L = 800 \, km,\, W_{X} = 0.1 \, GHz,\, W_{d} = 1\, GHz$. The corresponding dimensionless parameters of dispersion are $\beta L W_{X}^{2} = 0.00016$ and $\beta L W_{d}^{2} = 0.016$.
The contributions of $\beta L W_{d}^{2}I_{d}(\tilde{\gamma})$ and $\beta L W_{X}^{2}I_{X}(\tilde{\gamma})$ are shown in the fig. 3. We also show the first-order correction $\Delta I_{P_{X}}$ in the fig. 4.

\begin{figure}[hbt!]
	\centering
	\resizebox{.7\linewidth}{!}{\includegraphics{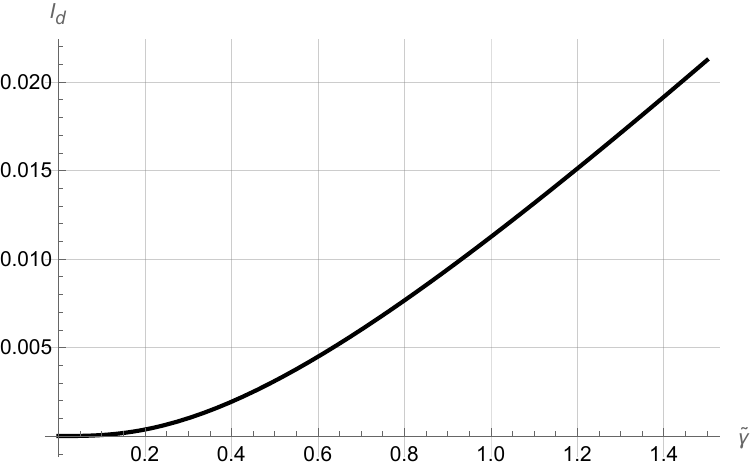}}
	\caption{The function $I_{d}(\tilde{\gamma})$ defined in the eq. \eqref{Iddefinition}.} 
	\label{fig:2}
\end{figure}

\begin{figure}[hbt!]
	\centering
	\resizebox{.7\linewidth}{!}{\includegraphics{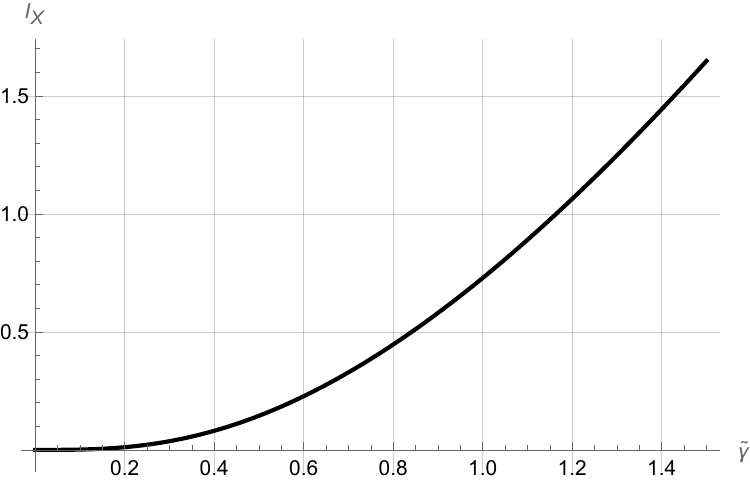}}
	\caption{The function $I_{X}(\tilde{\gamma})$ defined in the eq. \eqref{IXdefinition}.} 
	\label{fig:2}
\end{figure}

\begin{figure}[hbt!]
	\centering
	\resizebox{.7\linewidth}{!}{\includegraphics{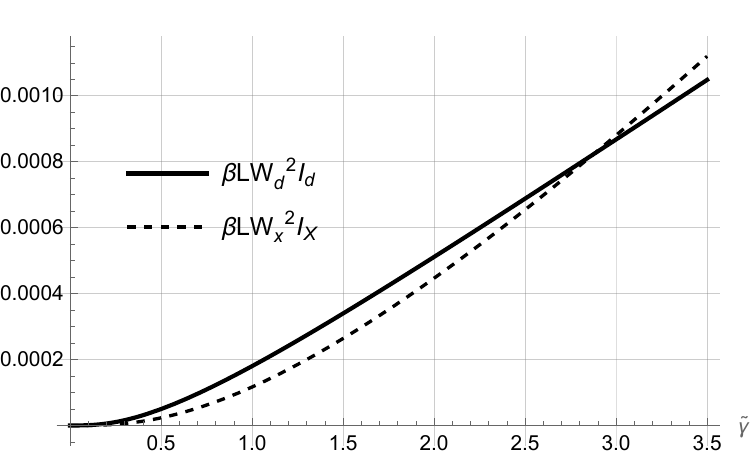}}
	\caption{Two contributions to the mutual information divided by $M_{d}$. The solid line corresponds to the positive contribution which is proportional to the function $I_{d}(\tilde{\gamma})$, and the dashed line corresponds to the negative contribution which is proportional to the function $I_{X}(\tilde{\gamma})$.} 
	\label{fig:2}
\end{figure}

\begin{figure}[hbt!]
	\centering
	\resizebox{.7\linewidth}{!}{\includegraphics{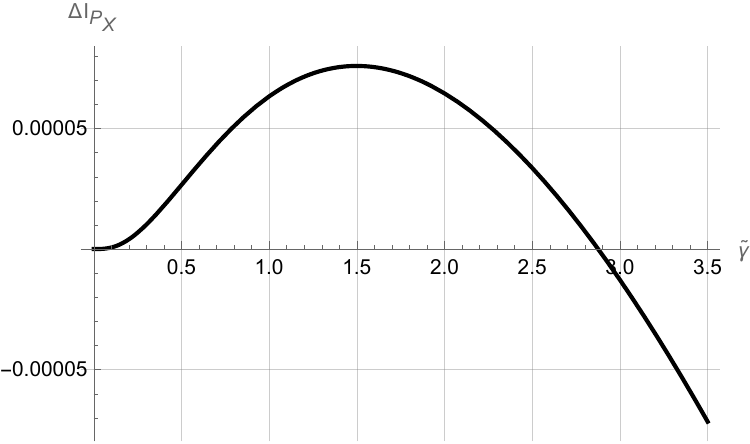}}
	\caption{The first-order correction in $\beta$ to the mutual information divided by $M_{d}$ for $\beta L W_{X}^{2} = 0.00016$ and $\beta L W_{d}^{2} = 0.016$.} 
	\label{fig:2}
\end{figure}

In the fig. 4 we observe that the correction reaches a maximum value at a point around $\tilde{\gamma} = 1.5$. However, such a point only exists when the ratio $W_{d}/W_{X}$ belongs to the specific region, which can be obtained from a close inspection of the function $G(\tilde{\gamma}, v)$. The maximum point disappears when the value of $v$ becomes smaller than the number $r_{1}$, which is defined as: 
\begin{eqnarray}\label{5r1definition}
    r_{1} = \sqrt{\left(\frac{I_{X}(y)}{y^{2}}\right)'\bigg/ \left(\frac{I_{d}(y)}{y^{3}}\right)}\bigg |_{y=0}
\end{eqnarray}
and which is roughly equal to $4.84$. The maximum point also disappears when $v$ becomes larger than $r_{2} = \sqrt{186}$. We show all of the three cases in the fig. 5. 
\begin{figure}[hbt!]
	\centering
	\resizebox{.7\linewidth}{!}{\includegraphics{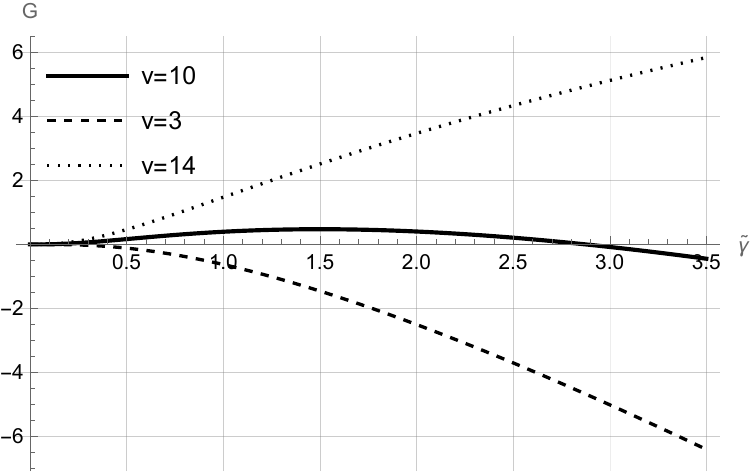}}
	\caption{Behaviour of $G(\tilde{\gamma}, v)$ as a function of $\tilde{\gamma}$ for different values of $v$. The solid, dashed and dotted lines correspond to the values $v=10$, $v=3$, $v=14$, correspondingly.} 
	\label{fig:2}
\end{figure}

Eventually, we see that there are three different regimes of the behaviour of the function $G(\tilde{\gamma}, v)$ depending on the value of $v=W_{d}/W_{X}$. When the ratio $W_{d}/W_{X}$ is large (that is, $W_{d}/W_{X}>\sqrt{186}$), the function $G(\gamma LP, v)$ always grows if $\tilde{\gamma}$ is increasing. Our interpretation is that for large bandwidth of the detector the spectral broadening of the signal (which is due to the nonlinearity of the channel) is not significant enough to drive the harmonics of the output signal out of the region $[-W_{d}/2,W_{d}/2]$.

If the ratio $W_{d}/W_{X}$ belongs to the region $[r_{1},r_{2}]$, the behaviour is a bit more complex. At first the function $G(\tilde{\gamma}, v)$ grows as we increase the dimensionless nonlinearity parameter $\tilde{\gamma}$, but after a certain point the spectral broadening drives harmonics of the signal out of the region $[-W_{d}/2,W_{d}/2]$, and we lose the information which is contained in the harmonics of frequencies $|\omega|>W_{d}/2$.

And finally, there exist values of the detector bandwidth for which the impact of the nonlinearity is too significant, so there is not a single value of the nonlinearity parameter $\tilde{\gamma}$ for which our detecting procedure succeeds in gathering all of the harmonics of the output signal.

We remind that our results are valid for the intermediate power range, i.e. for such $P$ that $N \ll P \ll (N\gamma^{2}L^{2})^{-1}$.

\section{Conclusion}

In the present paper we consider a channel described by the NLSE with additive Gaussian noise, arbitrary Kerr nonlinearity and small dimensionless second dispersion parameter. We also consider the signal-to-noise ratio (SNR) to be large. Using the path-integral approach and the saddle-point approximation we derive the conditional PDF $P[Y|X]$ in the leading order in $1/\textrm{SNR}$. The expression for the PDF $P[Y|X]$ obtained in this way can be expanded in powers of the second dispersion $\beta$. We thus obtain the first-order correction in $\beta$ to the conditional PDF in the eq. \eqref{Pdtildet}. To account for the finite bandwidth of the receiver $W_{d}$ we integrate $P[Y|X]$ over high-frequency Fourier components of the output signal $Y$, which are not distinguished by the receiver, and denote the conditional PDF for the observable components of the output signal as $P_{d}[Y|X]$. For the perturbative expansion of $P_{d}[Y|X]$ to be valid, the bandwidth of the input signal $W_{X}$, the bandwidth of the receiver $W_{d}$ and the second dispersion $\beta$ must obey the conditions $\beta L W_{X}^{2} \ll 1$, $\beta L W_{d}^{2} \ll 1$ and $W_{X} \lesssim W_{d}$. Using the conditional PDF $P_{d}[Y|X]$ we calculate the conditional entropy, probability density functional for the output signal, entropy of the output signal and mutual information. 

The expression for the mutual information can be used to find the optimal distribution of the input signal or to evaluate the transmission rate for some particular choice of the distribution of the input signal. We calculate the transmission rate for the Gaussian input signals with fixed average power and fixed average bandwidth $W_{X}$. We choose the input signal distribution in the form \eqref{PXosc}, which is essentially the case of two independent quantum harmonic oscillators evolving with imaginary time. Using well-known expressions for the correlators, we evaluate the averages in the expression for the mutual information \eqref{mutualinformation}. 

The behaviour of the first-order correction in dimensionless dispersion parameter to the mutual information, which we considered as a function of the dimensionless nonlinearity parameter $\gamma LP$, turned to have three different regimes depending on the value of the ratio $W_{d}/W_{X}$. In the first case, when $W_{d}/W_{X} > \sqrt{186}$, the correction is positive for all the values of $\gamma LP$. In the second case, when $W_{d}/W_{X}$ is less than the number $r_{1} \approx 4.84$, which is defined in the eq. \eqref{5r1definition}, the correction is negative for all the values of $\gamma LP$. The third regime is achieved when $r_{1} < W_{d}/W_{X} < \sqrt{186}$. In the third case the correction to the mutual information grows for small values of $\gamma LP$, then reaches a maximum, and then falls for all values of $\gamma LP$ larger than the value of the maximum point. 

\subsection*{Acknowledgements}
This work was supported by the Ministry of Science and Higher Education of the Russian Federation.

\section{Appendix A}
In this section we present explicitly the cumbersome expressions one may meet trying to find the function $\Psi_{cl}$ which makes the action functional from the eq. \eqref{3action} stationary. We recall that we expanded the function $\Psi_{cl}$ around the solution to the NLSE without noise:
\begin{eqnarray}
    \Psi_{cl} = \Phi + \kappa \exp \left\{i \mu \frac{z}{L} + i\phi\right\}, 
\end{eqnarray}
where $\mu = \gamma L |X|^{2}$ and $X = |X|e^{i\phi}$. We obtained the following problem for the first-order correction in $\beta$ to the function $\kappa=\kappa_0+\kappa_1+\ldots$, where $\kappa_0$ is defined in eqs.~\eqref{kappa0eqs}:
\begin{eqnarray}\label{Aboundaryproblem}
\begin{cases}
        & \partial_{z}^{2}\kappa_{1} - 2i\gamma \rho^{2}\kappa_{1} - 4(\gamma \rho^{2})^{2}\textrm{Re}[\kappa_{1}] = F(z,t), \\
        &\kappa_{1}(z=0,t) = 0, \\
        &\kappa_{1}(z=L,t) = 0,
    \end{cases}
\end{eqnarray}
where $\rho = |X|$. The cumbersome function $F(z,t)$ in the right-hand side reads: 
\begin{eqnarray}\label{AfunctionF}
    &&F(z,t) = -\partial_{z}\delta L_{1} - 2\gamma\rho^{2}\textrm{Im}[\delta L_{1}] - i\beta e^{-i(z\gamma\rho^2+\phi)}\partial^2_{t}(e^{i(z\gamma\rho^2+\phi)}L_{0}) + \nonumber\\
    &&2i\gamma\rho^{2}\Big(2L_{0}\textrm{Re}[\Phi_{1}e^{-i(z\gamma\rho^2+\phi)}]-\overline{L}_{0}\Phi_{1}e^{-i(z\gamma\rho^2+\phi)}\Big),\nonumber\\
    &&\delta L_{1} = i\beta e^{-i(z\gamma\rho^2+\phi)}\partial^2_{t}(e^{i(z\gamma\rho^2+\phi)}\kappa_{0}) -\nonumber\\
    &&2i\gamma\rho\Big(2\kappa_{0}\textrm{Re}[\Phi_{1}e^{-i(z\gamma\rho^2+\phi)}] + \overline{\kappa}_{0}\Phi_{1}e^{-i(z\gamma\rho^2+\phi)}\Big),\nonumber\\
    &&L_{0} = \partial_{z}\kappa_{0} - 2i\gamma\rho^{2}\textrm{Re}[\kappa_{0}].
\end{eqnarray}

The solution to the boundary problem \eqref{Aboundaryproblem} can be written using the Green's function method in the following form:
\begin{align}
    \kappa_{1}(z,t) = -\frac{1}{L}\int_{0}^{L}dz'\left[G_{F}(z,t;z')F(z',t) + G_{\bar{F}}(z,t;z')\bar{F}(z',t)\right],
\end{align}
where the functions $G_{F}(z,t;z')$ and $G_{\bar{F}}(z,t;z')$ read:
\begin{align}\label{AGFfunc}
    &G_{F}(z,t;z') = \frac{1}{3(\mu^{2}+3)L^{6}}\Bigg[z(L-z')\theta(z\leq z')\Big\{(\mu^{2}+3)L^{2}(3L^{2}+3i\mu L z - \nonumber\\ 
    &\mu^{2}z^{2})+\mu L z'\big(-3i(\mu^{2}-i\mu+3)L^{2}+3\mu L z(\mu^{2}-i\mu+3)-\mu^{2}(\mu-3i)z^{2}\big)+\nonumber\\
    &\mu^{2}(z')^{2}\big(3i(\mu+i)L^{2}-
    3\mu L z(\mu + i)+2\mu^{2}z^{2}\big)\Big\}+z'(L-z)\theta(z' < z)\nonumber\\
    &\Big\{3L^{2}\big((\mu^{2}+3)L^{2}+i\mu L z(\mu^{2}+i\mu + 3)-(1+i\mu)\mu^{2}z^{2}\big)+\nonumber\\
    &3\mu L z'\big(-i(\mu^{2}+3)L^{2}+L^{2}\mu z(\mu^{2}+i\mu +3)-\mu^{2}z^{2}(\mu-i)\big)-\nonumber\\
    &\mu^{2}(z')^{2}\big((\mu^{2}+3)L^{2}+\mu L z(\mu+3i)-2\mu^{2}z^{2}\big)\Big\}\Bigg],
\end{align}
\begin{align}\label{AGFbarfunc}
    &G_{\bar{F}}(z,t;z') = \frac{\mu^{2}}{3(\mu^{2}+3)L^{6}}\Bigg[z'(L-z)\theta(z' < z)\Big\{(z')^{2}\big((\mu^{2}+3)L^{2}+\mu L z(\mu+3i)-\nonumber\\
    &2\mu^{2}z^{2}\big)+3L^{2}z\big(iL(\mu+2i)-i\mu z+z\big)+3\mu L z z'\big((\mu+i)z-(\mu+2i)L\big)\Big\}+\nonumber\\
    &z(L-z')\theta(z \leq z')\Big\{(\mu^{2}+3)L^{2}z^{2}+(z')^{2}\big(3L^{2}(1-i\mu)+3\mu L z(\mu+i)-2\mu^{2} z^{2}\big)+\nonumber\\
    &L z'\big(3iL^{2}(\mu+2i)-3\mu L z(\mu + 2i)+\mu z^{2}(\mu + 3i)\big)\Big\}\Bigg].
\end{align}
The function $\theta(x < y)$ in eqs. \eqref{AGFfunc} and \eqref{AGFbarfunc} is the Heaviside step function. 

After finding $\kappa_{1}$ we can calculate $S_{cl} \equiv S[\Psi_{cl}]$. We write the next-to-leading order correction in $\beta$ to the classical action $S_{cl}$ in the following form: 
\begin{eqnarray}
    &\!S_{cl}^{(1)} = \beta \int_{T} dt \Big\{a_{1}(t)x^{2} + a_{2}(t)y^{2} + a_{3}(t)xy + a_{4}(t)x\dot{x} + a_{5}(t)y\dot{y} +\nonumber\\
    &\!a_{6}(t)x\dot{y} + a_{7}(t)\dot{x}y + a_{8}(t)x\ddot{x} + a_{9}(t)y\ddot{y} + a_{10}(t)x\ddot{y} + a_{11}(t)\ddot{x}y\Big\},
\end{eqnarray}
where the coefficients $a_{i}(t)$ are functions of time which we express through the input signal $X(t)$:
\begin{eqnarray}\label{a1}
    &&a_{1}(t) = \frac{1}{15\mu(3+\mu^{2})^{4}} (90 \mu ^9 \ddot{\phi }+30 \mu ^9 \ddot{\mu }+90 \mu ^8 \dot{\phi }^2+180 \dot{\mu } \mu ^8 \dot{\phi }+ \nonumber\\
    &&62 \dot{\mu }^2 \mu ^8+1035 \mu ^7 \ddot{\phi } +288 \mu ^7 \ddot{\mu }+810 \mu ^6 \dot{\phi }^2+2070 \dot{\mu } \mu ^6 \dot{\phi }+780 \dot{\mu }^2 \mu ^6+\nonumber\\
    &&3645 \mu ^5 \ddot{\phi }+999 \mu ^5 \ddot{\mu }+2430 \mu ^4 \dot{\phi }^2+7560 \dot{\mu } \mu ^4 \dot{\phi }+2403 \dot{\mu }^2 \mu ^4+3645 \mu ^3 \ddot{\phi }+ \nonumber\\
    &&810 \mu ^3 \ddot{\mu } +2430 \mu ^2 \dot{\phi }^2+8910 \dot{\mu } \mu ^2 \dot{\phi }+4590 \dot{\mu }^2 \mu ^2-1215 \mu  \ddot{\phi }-1215 \mu  \ddot{\mu }+\nonumber\\
    &&405 \dot{\mu }^2),
\end{eqnarray}
\begin{eqnarray}\label{a2}
    &&a_{2}(t) = -\frac{1}{5\mu(3+\mu^{2})^{4}} (30 \mu ^9 \ddot{\phi }+10 \mu ^9 \ddot{\mu }+30 \mu ^8 \dot{\phi }^2+60 \dot{\mu } \mu ^8 \dot{\phi }+\nonumber\\
    &&20 \dot{\mu }^2 \mu ^8+315 \mu ^7 \ddot{\phi }+96 \mu ^7 \ddot{\mu }+270 \mu ^6 \dot{\phi }^2+570\dot{\mu } \mu ^6 \dot{\phi }+198 \dot{\mu }^2 \mu ^6+\nonumber\\
    &&1125 \mu ^5 \ddot{\phi }+315 \mu ^5 \ddot{\mu }+810 \mu ^4 \dot{\phi }^2+1800 \dot{\mu } \mu ^4 \dot{\phi }+639 \dot{\mu }^2 \mu ^4+1485 \mu ^3 \ddot{\phi }+\nonumber\\
    &&306 \mu ^3 \ddot{\mu }+810 \mu ^2 \dot{\phi }^2+1890 \dot{\mu } \mu ^2 \dot{\phi }+1008 \dot{\mu }^2 \mu ^2+405 \mu  \ddot{\phi }-135 \mu  \ddot{\mu }+\nonumber\\
    &&135 \dot{\mu }^2),
\end{eqnarray}
\begin{eqnarray}\label{a3}
    &&a_{3}(t) = \frac{1}{15(3+\mu^{2})^{4}}(120 \mu ^9 \ddot{\phi }+40 \mu ^9 \ddot{\mu }+120 \mu ^8 \dot{\phi }^2+240 \dot{\mu } \mu ^8 \dot{\phi }+ \nonumber\\
    && 80 \dot{\mu }^2 \mu ^8+1200 \mu ^7 \ddot{\phi }+363 \mu ^7 \ddot{\mu }+1080 \mu ^6 \dot{\phi}^2+2280 \dot{\mu } \mu ^6 \dot{\phi }+792 \dot{\mu }^2 \mu ^6+\nonumber\\
    &&3780 \mu ^5 \ddot{\phi } + 1017 \mu ^5 \ddot{\mu }+3240 \mu ^4 \dot{\phi }^2+6840 \dot{\mu } \mu ^4 \dot{\phi }+2412 \dot{\mu }^2 \mu^4+3240 \mu ^3 \ddot{\phi }+\nonumber\\
    &&189 \mu ^3 \ddot{\mu }+3240 \mu ^2 \dot{\phi }^2+5400 \dot{\mu } \mu ^2 \dot{\phi }+3456 \dot{\mu }^2 \mu ^2-1620 \mu  \ddot{\phi }-3240 \dot{\mu } \dot{\phi }- \nonumber\\
    &&2025 \mu \ddot{\mu } -1620 \dot{\mu }^2),
\end{eqnarray}
\begin{eqnarray}\label{a4}
    &&a_{4}(t) = - \frac{2}{15(3+\mu^{2})^{3}} (30 \mu ^6 \dot{\phi }+14 \dot{\mu } \mu ^6+225 \mu ^4 \dot{\phi }+99 \dot{\mu } \mu ^4+540 \mu ^2 \dot{\phi }+\nonumber\\
    &&135 \dot{\mu }+405 \dot{\phi }),
\end{eqnarray}
\begin{eqnarray}\label{a5}
    &&a_{5}(t) = -\frac{6}{5(3+\mu^{2})^{3}} (5 \mu ^4 \dot{\phi }+3 \dot{\mu } \mu ^4+30 \mu ^2 \dot{\phi }+22 \dot{\mu } \mu ^2+15 \dot{\mu }+\nonumber\\
    &&45 \dot{\phi }),
\end{eqnarray}
\begin{eqnarray}\label{a6}
    &&a_{6}(t) = \frac{\mu}{5(3+\mu^{2})^{3}} (20 \mu ^4 \dot{\phi }+13 \dot{\mu } \mu ^4+120 \mu ^2 \dot{\phi }+102 \dot{\mu } \mu ^2 + \nonumber\\
    &&45 \dot{\mu }+180 \dot{\phi }),
\end{eqnarray}
\begin{eqnarray}\label{a7}
    &&a_{7}(t) = \frac{\mu}{5(3+\mu^{2})^{3}} (20 \mu ^4 \dot{\phi }+11 \dot{\mu } \mu ^4+120 \mu ^2 \dot{\phi }+66 \dot{\mu } \mu ^2-45 \dot{\mu } + \nonumber\\
    &&180 \dot{\phi }),
\end{eqnarray}
\begin{eqnarray}\label{a8}
    &&a_{8}(t) = \frac{2\mu}{15(3+\mu^{2})^{2}} (-45 + 15\mu^{2} + 4 \mu^{4}),
\end{eqnarray}
\begin{eqnarray}\label{a9}
    &&a_{9}(t) = \frac{6\mu}{5(3+\mu^{2})^{2}} (5+\mu^{2}),
\end{eqnarray}
\begin{eqnarray}\label{a10}
    &&a_{10}(t) = -\frac{1}{5(3+\mu^{2})^{2}} (45 + 60\mu^{2} + 11\mu^{4}),
\end{eqnarray}
\begin{eqnarray}\label{a11}
    &&a_{11}(t) = -\frac{1}{5(3+\mu^{2})^{2}} (-45 + \mu^{4}).
\end{eqnarray}

\section{Appendix B}
This section contains detailed calculation of the conditional PDF $\tilde{P}_{d}[Y_{d}|X]$ from the eq. \eqref{Pdtildet}. We start from the PDF from the eq. \eqref{P[Y|X]quas}, which contains information about all of the harmonics of the output signal $Y(t)$. Up to the first order in $\beta$ it reads: 
\begin{eqnarray}\label{BP[Y|X]initial}
    P[Y|X] = \Lambda \exp \left\{-\frac{S_{cl}^{(0)}[\tilde{Y}]}{Q}\right\}\left(1-\frac{S_{cl}^{(1)}[\tilde{Y}]}{Q}\right),
\end{eqnarray}
where we write the classical action $S_{cl} \equiv S[\Psi_{cl}]$ as a functional with the variable $\tilde{Y}(t) = y_{1}(t)+iy_{2}(t)$ defined in the eq. \eqref{subsYtildeY}. The leading contribution in $\beta$ to the action is diagonal in $y_{1,2}(t)$:
\begin{eqnarray}
    S_{cl}^{(0)}[\tilde{Y}] = \frac{1}{L}\int_{T}dt\left[y_{1}^{2}(t)+y_{2}^{2}(t)\right],
\end{eqnarray}
and the first-order correction can be written as: 
\begin{eqnarray}
    &&S_{cl}^{(1)}[\tilde{Y}] = \beta \int_{T} dt \{b_{1}y_{1}^{2} + b_{2}y_{2}^{2} + b_{3}y_{1}y_{2} + b_{4}y_{1}\dot{y}_{1} + b_{5}y_{2}\dot{y}_{2} + \nonumber\\
    &&b_{6}y_{1}\dot{y}_{2} + b_{7}\dot{y}_{1}y_{2} + b_{8}y_{1}\ddot{y}_{1} + b_{9}y_{2}\ddot{y}_{2} + b_{10}y_{1}\ddot{y}_{2} + b_{11}\ddot{y}_{1}y_{2}\},
\end{eqnarray}
where the functions $b_{i}(t)$ can be expressed through $a_{i}(t)$ (see eqs. \eqref{a1}-\eqref{a11}) and $A_{ij}(t)$ (see eq. \eqref{Aij}):
\begin{eqnarray}\label{b1}
    &&\mkern-36mu b_{1}(t) = a_4 A_{11} \dot{A}_{11}+a_6 A_{11} \dot{A}_{21}+a_7 A_{21} \dot{A}_{11}+a_5 A_{21} \dot{A}_{21}+a_8 A_{11} \ddot{A}_{11}+\nonumber\\
    &&\mkern-36mu a_{10} A_{11} \ddot{A}_{21}+a_{11} A_{21} \ddot{A}_{11}+a_9 A_{21} \ddot{A}_{21}+a_1 A_{11}^2+a_3 A_{21} A_{11}+a_2
   A_{21}^2,
\end{eqnarray}
\begin{eqnarray}\label{b2}
    &&\mkern-36mu b_{2}(t) = a_4 A_{12} \dot{A}_{12}+a_6 A_{12} \dot{A}_{22}+a_7 A_{22} \dot{A}_{12}+a_5 A_{22} \dot{A}_{22}+a_8 A_{12} \ddot{A}_{12}+\nonumber\\
    &&\mkern-36mua_{10} A_{12} \ddot{A}_{22}+a_{11} A_{22} \ddot{A}_{12}+a_9 A_{22} \ddot{A}_{22}+a_1 A_{12}^2+a_3 A_{22} A_{12}+a_2 A_{22}^2,
\end{eqnarray}
\begin{eqnarray}\label{b3}
    &&\mkern-36mu b_{3}(t) = a_4 \left(A_{12} \dot{A}_{11} + A_{11} \dot{A}_{12}\right)+a_7 \left(A_{22} \dot{A}_{11}+ A_{21} \dot{A}_{12}\right)+\nonumber\\
    &&\mkern-36mu a_5 \left(A_{22} \dot{A}_{21} + A_{21} \dot{A}_{22}\right)+a_6 \left(A_{11} \dot{A}_{22} + A_{12} \dot{A}_{21}\right)+a_{10}\left( A_{12} \ddot{A}_{21}+ A_{11} \ddot{A}_{22}\right)+\nonumber\\
    &&\mkern-36mu a_{11}\left(A_{22} \ddot{A}_{11}+ A_{21} \ddot{A}_{12}\right)+a_8 \left( A_{11} \ddot{A}_{12} + A_{12} \ddot{A}_{11}\right)+a_9 \left( A_{22} \ddot{A}_{21}+ A_{21} \ddot{A}_{22}\right)+\nonumber\\
    &&\mkern-36mu a_3 \left(A_{21} A_{12}+A_{11} A_{22}\right)+2 a_2 A_{21} A_{22} + 2 a_1 A_{11} A_{12},
\end{eqnarray}
\begin{eqnarray}\label{b4}
    &&\mkern-36mu b_{4}(t) = 2 a_8 A_{11} \dot{A}_{11}+2 a_{10} A_{11} \dot{A}_{21}+2 a_{11} A_{21} \dot{A}_{11}+2 a_9 A_{21} \dot{A}_{21}+a_4 A_{11}^2+\nonumber\\
    &&\mkern-36mua_6 A_{21} A_{11}+a_7 A_{21} A_{11}+a_5 A_{21}^2,
\end{eqnarray}
\begin{eqnarray}\label{b5}
    &&\mkern-36mu b_{5}(t) = 2 a_8 A_{12} \dot{A}_{12}+2 a_{10} A_{12} \dot{A}_{22}+2 a_{11} A_{22} \dot{A}_{12}+2 a_9 A_{22} \dot{A}_{22}+a_4 A_{12}^2+\nonumber\\
    &&\mkern-36mu a_6 A_{22} A_{12}+a_7 A_{22} A_{12}+a_5 A_{22}^2,
\end{eqnarray}
\begin{eqnarray}\label{b6}
    &&\mkern-36mu b_{6}(t) = 2 a_8 A_{11} \dot{A}_{12}+2 a_{11} A_{21} \dot{A}_{12}+2 a_{10} A_{11} \dot{A}_{22}+2 a_9 A_{21} \dot{A}_{22}+a_4 A_{11} A_{12}+\nonumber\\
    &&\mkern-36mu a_7 A_{21} A_{12}+a_6 A_{11} A_{22}+a_5 A_{21} A_{22},
\end{eqnarray}
\begin{eqnarray}\label{b7}
    &&\mkern-36mu b_{7}(t) = 2 a_8 A_{12} \dot{A}_{11}+2 a_{11} A_{22} \dot{A}_{11}+2 a_{10} A_{12} \dot{A}_{21}+2 a_9 A_{22} \dot{A}_{21}+a_4 A_{11} A_{12}+\nonumber\\
    &&\mkern-36mu a_7 A_{11} A_{22}+a_6 A_{12} A_{21}+a_5 A_{21} A_{22},
\end{eqnarray}
\begin{eqnarray}\label{b8}
    &&\mkern-36mu b_{8}(t) = a_8 A_{11}^2+a_{10} A_{21} A_{11}+a_{11} A_{21} A_{11}+a_9 A_{21}^2,
\end{eqnarray}
\begin{eqnarray}\label{b9}
    &&\mkern-36mu b_{9}(t) = a_8 A_{12}^2+a_{10} A_{22} A_{12}+a_{11} A_{22} A_{12}+a_9 A_{22}^2,
\end{eqnarray}
\begin{eqnarray}\label{b10}
    &&\mkern-36mu b_{10}(t) = a_8 A_{11} A_{12}+a_{11} A_{21} A_{12}+a_{10} A_{11} A_{22}+a_9 A_{21} A_{22},
\end{eqnarray}
\begin{eqnarray}\label{b11}
    &&\mkern-36mu b_{11}(t) = a_8 A_{11} A_{12}+a_{10} A_{21} A_{12}+a_{11} A_{11} A_{22}+a_9 A_{21} A_{22}.
\end{eqnarray}
To perform the integration over the high-frequency harmonics of the function $\tilde{Y}(t)$ we have to introduce a discrete time grid with the discretization step $\delta_t = 2\pi/W'$. The integrals over time are replaced with sums: 
\begin{eqnarray}
    \int_{T} dt \rightarrow \delta_t \sum_{n=0}^{M-1}.
\end{eqnarray}
As we will work with the Fourier harmonics of the function $\tilde{Y}(t)$, we also have to introduce the discrete Fourier transform (DFT) of a function $F(t)$ defined on the time grid:
\begin{eqnarray}
    &&F(t_{n}) = \sum_{n'=0}^{M-1} \hat{F}(\omega_{n'}) e^{\frac{2\pi i n n'}{M}},\label{BDFT1}\\
    &&\hat{F}(\omega_{n'}) = \frac{1}{M}\sum_{n=0}^{M-1} F(t_{n}) e^{-\frac{2\pi i n n'}{M}}.\label{BDFT2}
\end{eqnarray}
The discretization in the frequency domain is defined as: 
\begin{eqnarray}\label{Bfreqs}
    \begin{cases}
    \omega_{n} = 2\pi \delta_{\omega} n,\quad n = 0, ..., \left[\frac{M-1}{2}\right],\\
    \omega_{n} = -2\pi \delta_{\omega} (M - n),\quad n = \left[\frac{M-1}{2}\right] + 1, ..., M-1,
    \end{cases}
\end{eqnarray}
where $\delta_\omega = \frac{W'}{2\pi M}$ is the discretization step in the frequency domain, $W'$ is the bandwidth of the noise, and $[...]$ is the floor function. In our consideration the bandwidth $W'$ plays the role of the ultraviolet cutoff. 

The transition from the time domain to the frequency one is straightforward in $S_{cl}^{(0)}$:
\begin{eqnarray}
    \!\!\!S_{cl}^{(0)} = \frac{\delta_{t}}{L}\sum_{j=0}^{M-1} \left[y_{1}^{2}(t_{j}) + y_{2}^{2}(t_{j})\right] = \frac{M\delta_{t}}{L} \sum_{j'=0}^{M-1} \left[|\hat{y}_{1}(\omega_{j'})|^{2} + |\hat{y}_{1}(\omega_{j'})|^{2}\right],
\end{eqnarray}
due to the discrete version of the Parseval's theorem:
\begin{eqnarray}
    \frac{1}{M} \sum_{n=0}^{M-1} \Psi(t_{n}) \bar{\Phi}(t_{n}) = \sum_{n'=0}^{M-1} \hat{\Psi}(\omega_{n'}) \bar{\hat{\Phi}}(\omega_{n'}).
\end{eqnarray}
It is also worth mentioning that due to the functions $y_{i}(t)$ being real-valued, the Fourier-transform of $y_{i}(t)$ has the following property:
\begin{eqnarray}
    \bar{\hat{y}}_{i}(\omega) = \hat{y}_{i}(-\omega).
\end{eqnarray}

In the case of the first order correction $S_{cl}^{(1)}$, we will adopt the matrix notation: 
\begin{eqnarray}\label{BS_1matrix}
    &&S_{cl}^{(1)} = \beta M \delta_{t} \big\{\hat{y}_{1}^{\dag}\hat{b}_{1}\hat{y}_{1} + \hat{y}_{2}^{\dag}\hat{b}_{2}\hat{y}_{2} + \hat{y}_{1}^{\dag}\hat{b}_{3}\hat{y}_{2} + \hat{y}_{1}^{\dag}\hat{b}_{4}\hat{D}_{1}\hat{y}_{1} + \hat{y}_{2}^{\dag}\hat{b}_{5}\hat{D}_{1}\hat{y}_{2} + \nonumber\\
    && \hat{y}_{1}^{\dag}\hat{b}_{6}\hat{D}_{1}\hat{y}_{2} + \hat{y}_{2}^{\dag}\hat{b}_{7}\hat{D}_{1}\hat{y}_{1} + \hat{y}_{1}^{\dag}\hat{b}_{8}\hat{D}_{2}\hat{y}_{1} + 
    \hat{y}_{2}^{\dag}\hat{b}_{9}\hat{D}_{2}\hat{y}_{2} + 
    \hat{y}_{1}^{\dag}\hat{b}_{10}\hat{D}_{2}\hat{y}_{2} + \nonumber\\
    && \hat{y}_{2}^{\dag}\hat{b}_{11}\hat{D}_{2}\hat{y}_{1}\big\},
\end{eqnarray}
where the components of the matrices $\hat{b}_{i}$ and vectors $\hat{y}_{1,2}$ are defined as: 
\begin{eqnarray}
    &&(\hat{y}_{1,2})_{n} = \hat{y}_{1,2} (\omega_{n}),\\
    &&(\hat{b}_{i})_{n n'} = \hat{b}_{i}(\omega_{n} - \omega_{n'}).\label{Bbmatrix}
\end{eqnarray}
The matrices $\hat{D}_{1}$ and $\hat{D}_{2}$ correspond to the operators of the first and the second derivatives, respectively. We write it in that way to emphasize that for now we do not restrict our consideration to a specific choice of the discretization scheme for the time derivatives. The first time derivative of a function $g(t)$ after discretization in this notation reads: 
\begin{eqnarray}
    \dot{g}(t_{n}) = \sum_{m=0}^{M-1} (D_1)_{nm}g(t_{m}) = \sum_{m,n'=0}^{M-1} (\hat{D}_{1})_{n'm}\hat{g}(\omega_{m})e^{\frac{2\pi i n n'}{M}},
\end{eqnarray}
and the second derivative is defined in a similar way by replacing $D_{1}$ by $D_{2}$.

Now that we developed the notation we can move to integrating the PDF from the eq. \eqref{BP[Y|X]initial} over the high-frequency components of the vectors $\hat{y}_{1,2}$. According to the discretization of frequencies defined by the eq. \eqref{Bfreqs}, the vectors $\hat{y}_{1,2}$ can be divided into three parts:
\begin{eqnarray}\label{Byvector}
    \hat{y}_{1,2} = 
    \begin{pmatrix}
    \vec{\varphi}_{+}^{1,2}\\
    \vec{0}\\
    \vec{\varphi}_{-}^{1,2}
    \end{pmatrix}
    + 
    \begin{pmatrix}
    \vec{0}\\
    \vec{L}_{1,2}\\
    \vec{0}
    \end{pmatrix}.
\end{eqnarray}
The vectors $\vec{L}_{1,2}$ consist of the high-frequency harmonics ($|\omega|>\frac{W_{d}}{2}$) which we wish to get rid of by the integration \eqref{P_domegatilde}. On the other hand, the vectors $\vec{\phi}_{\pm}^{1,2}$ consist of the observable harmonics with low frequencies ($|\omega|<\frac{W_{d}}{2}$). Due to the diagonal form of the functional $S_{cl}^{(0)}$: 
\begin{eqnarray}
    S_{cl}^{(0)} = \frac{M\delta_{t}}{L}\left[\vec{\chi}_{1}^{\dag}\vec{\chi}_{1} + \vec{\chi}_{2}^{\dag}\vec{\chi}_{2}\right] + \frac{M\delta_{t}}{L}\left[\vec{L}_{1}^{\dag}\vec{L}_{1} + \vec{L}_{2}^{\dag}\vec{L}_{2}\right],
\end{eqnarray}
the integration over the components of the vectors $\vec{L}_{1,2}$ can be done using the following formulas for Gaussian integrals:
\begin{eqnarray}\label{Lformula1}
    \!\!\!\int D\vec{L}_{1}D\vec{L}_{2} \exp\left\{-\frac{M\delta_{t}}{QL}\left[\vec{L}_{1}^{\dag}\vec{L}_{1} + \vec{L}_{2}^{\dag}\vec{L}_{2}\right]\right\} = \left(\frac{\pi Q L}{M \delta_{t}}\right)^{M-M_{d}}, 
\end{eqnarray}
\begin{eqnarray}\label{Lformula2}
  &&\int D\vec{L}_{1}D\vec{L}_{2} \vec{L}_{\alpha}^{\dag} \hat{A} \vec{L}_{\beta} \exp\left\{-\frac{M\delta_{t}}{QL}\left[\vec{L}_{1}^{\dag}\vec{L}_{1} + \vec{L}_{2}^{\dag}\vec{L}_{2}\right]\right\} =  \nonumber\\
  &&= \left(\frac{\pi Q L}{M \delta_{t}}\right)^{M-M_{d}}\frac{QL}{M\delta_{t}}Tr[\hat{A}] \delta_{\alpha \beta},
\end{eqnarray}
where in the integration measure $D\vec{L}_{i} = \prod_{n} d \textrm{Re} \hat{y}_{i}(\omega_{n}) d \textrm{Im} \hat{y}_{i}(\omega_{n})$ the product is taken over such frequencies that $|\omega_{n}|>W_{d}/2$. We denoted by $M_{d}$ the amount of the frequencies $\omega_{n}$ which correspond to the observable low-frequency harmonics. These harmonics are contained in the vectors $\vec{\chi}_{1,2}$:
\begin{eqnarray}
    \vec{\chi}_{1, 2} = 
    \begin{pmatrix}
    \vec{\varphi}_{+}^{1,2}\\
    \vec{\varphi}_{-}^{1,2}
    \end{pmatrix}.
\end{eqnarray}
The result of the integration can be written as follows: 
\begin{eqnarray}\label{BPdtildeomega0}
    &&\tilde{P}_{d}[\tilde{Y}_{d}|X] \equiv \int D\vec{L}_{1}D\vec{L}_{2} \tilde{P}[\tilde{Y}(t)|X] = \nonumber \\
    &&\Lambda \left(\frac{\pi Q L}{M \delta_{t}}\right)^{M-M_{d}} \exp\left\{-\frac{M\delta_{t}}{QL}\left[\vec{\chi}_{1}^{\dag}\vec{\chi}_{1} + \vec{\chi}_{2}^{\dag}\vec{\chi}_{2}\right]\right\}\times\nonumber\\
    && \Big(1 - \frac{\beta M \delta_{t}}{Q} \Big[\vec{\chi}_{1}^{\dag}\hat{b}_{1}^{d}\vec{\chi}_{1} + \vec{\chi}_{2}^{\dag}\hat{b}_{2}^{d}\vec{\chi}_{2} + \vec{\chi}_{1}^{\dag}\hat{b}_{3}^{d}\vec{\chi}_{2} + \vec{\chi}_{1}^{\dag}\hat{b}_{4}^{d}\hat{d}_{1}\vec{\chi}_{1} + \nonumber\\
    && \vec{\chi}_{2}^{\dag}\hat{b}_{5}^{d}\hat{d}_{1}\vec{\chi}_{2} + 
    \vec{\chi}_{1}^{\dag}\hat{b}_{6}^{d}\hat{d}_{1}\vec{\chi}_{2} + 
    \vec{\chi}_{2}^{\dag}\hat{b}_{7}^{d}\hat{d}_{1}\vec{\chi}_{1} + 
    \vec{\chi}_{1}^{\dag}\hat{b}_{8}^{d}\hat{d}_{2}\vec{\chi}_{1} + 
    \vec{\chi}_{2}^{\dag}\hat{b}_{9}^{d}\hat{d}_{2}\vec{\chi}_{2} + \nonumber\\
    &&\vec{\chi}_{1}^{\dag}\hat{b}_{10}^{d}\hat{d}_{2}\vec{\chi}_{2} + 
    \vec{\chi}_{2}^{\dag}\hat{b}_{11}^{d}\hat{d}_{2}\vec{\chi}_{1}\Big] - \beta L  Tr\Big[\hat{b}_{1}^{L} + \hat{b}_{2}^{L} + \hat{b}_{4}^{L} \hat{D}_{1}^{L} + \nonumber\\
    && \hat{b}_{5}^{L} \hat{D}_{1}^{L} + \hat{b}_{8}^{L} \hat{D}_{2}^{L} + \hat{b}_{9}^{L} \hat{D}_{2}^{L} \Big]\Big).
\end{eqnarray}
In the eq. \eqref{BPdtildeomega0} we have introduced some more notation with matrices $\hat{b}_{i}^{d, L}, \hat{d}_{1,2}$ and $\hat{D}_{1,2}^{L}$ to describe what happens to the matrices $\hat{b}_{i}$ and $\hat{D}_{1,2}$ after we divide vectors $\hat{y}_{1,2}$ into low- and high-frequency parts according to the eq. \eqref{Byvector}. To explain the definition of the new matrices, let us first write $\hat{b}_{i}$ and $\hat{D}_{1,2}$ as the following block matrices:
\begin{align}
 &\hat{b}_{i} = 
 \begin{pmatrix}\label{Bblockmatrices1}
    (\hat{b}_{i})_{11} & (\hat{b}_{i})_{12} & (\hat{b}_{i})_{13} \\
    (\hat{b}_{i})_{21} & (\hat{b}_{i})_{22} & (\hat{b}_{i})_{23} \\
    (\hat{b}_{i})_{31} & (\hat{b}_{i})_{32} & (\hat{b}_{i})_{33} \\
 \end{pmatrix}, \\
 &\hat{D}_{1,2} = 
 \begin{pmatrix}\label{Bblockmatrices2}
    (\hat{D}_{1,2})_{11} & (\hat{D}_{1,2})_{12} & (\hat{D}_{1,2})_{13} \\
    (\hat{D}_{1,2})_{21} & (\hat{D}_{1,2})_{22} & (\hat{D}_{1,2})_{23} \\
    (\hat{D}_{1,2})_{31} & (\hat{D}_{1,2})_{32} & (\hat{D}_{1,2})_{33} \\
 \end{pmatrix}.
\end{align}
Using the block representation from the eqs. \eqref{Bblockmatrices1} and \eqref{Bblockmatrices2}, the scalar products which one meets computing $\tilde{P}_{d}[\tilde{Y}_{d}|X]$ can be written as follows:  
\begin{eqnarray}\label{Bsmallbddefinition}
    &&\begin{pmatrix}
        \vec{\varphi}_{+}^{1,2}, \vec{0}, \vec{\varphi}_{-}^{1,2}
    \end{pmatrix}^{\dag}
    \hat{b}_{i} \hat{D}_{1,2}
    \begin{pmatrix}
        \vec{\varphi}_{+}^{1,2}\\
        \vec{0}\\
        \vec{\varphi}_{-}^{1,2}
    \end{pmatrix}
    \approx \vec{\chi}_{1,2}^{\dagger}\hat{b}_{i}^{d}\hat{d}_{1,2}\vec{\chi}_{1,2}, 
\end{eqnarray}
\begin{eqnarray}
    \begin{pmatrix}
        \vec{0}, \vec{L}_{1,2}, \vec{0}
    \end{pmatrix}^{\dag}
    \hat{b}_{i} \hat{D}_{1,2}
    \begin{pmatrix}
        \vec{0}\\
        \vec{L}_{1,2}\\
        \vec{0}
    \end{pmatrix}
    = \vec{L}_{1,2}^{\dag} \hat{b}_{i}^{L} \hat{D}_{1,2}^{L} \vec{L}_{1,2},
\end{eqnarray}
where the new matrices are: 
\begin{align}
    &\hat{b}_{i}^{d} = 
    \begin{pmatrix} 
    (\hat{b}_{i})_{11} & (\hat{b}_{i})_{13}\\
    (\hat{b}_{i})_{31} & (\hat{b}_{i})_{33}
    \end{pmatrix}, \,\, 
    \hat{d}_{1,2} =  \begin{pmatrix} 
    (\hat{D}_{1,2})_{11} & (\hat{D}_{1,2})_{13}\\
    (\hat{D}_{1,2})_{31} & (\hat{D}_{1,2})_{33}
    \end{pmatrix},\nonumber\\
    &\hat{L}_{i}^{d} = (\hat{b}_{i})_{22}, \,\, 
    \hat{D}_{1,2}^{L} =  (\hat{D}_{1,2})_{22}.
\end{align}
Thus, the matrices $\hat{d}_{1,2}$ and $\hat{D}_{1,2}^{L}$ correspond to the low- and high-frequency parts of the derivative operators $\hat{D}_{1,2}$ respectively. The same goes for the matrices $\hat{b}_{i}^{L}$ and $\hat{b}_{i}^{d}$ - the superscript $L$ means that the matrix $\hat{b}_{i}^{L}$ consists of such components $(\hat{b}_{i})_{nm}$ that the corresponding frequencies $\omega_{n,m}$ are large: $|\omega_{n}|>W_{d}/2$ and $|\omega_{m}|>W_{d}/2$. In turn, the matrix $\hat{b}_{i}^{d}$ consists of low-frequency components of $(\hat{b}_{i})_{nm}$. We wrote approximate equality in the eq. \eqref{Bsmallbddefinition} because in that step we neglected contribution of the blocks $(\hat{b}_{i})_{12}$ and $(\hat{b}_{i})_{32}$. These blocks contain high-frequency Fourier harmonics of the functions $b_{i}(t)$. But we consider slow functions $b_{i}(t)$ (which means that $\beta L W_{X}^{2} \ll 1$), so the high-frequency harmonics are much smaller than the low-frequency ones.

What is left now is finding the normalization factor $\Lambda$, which we will do by using the normalization condition for the PDF $\tilde{P}_{d}[\tilde{Y}_{d}|X]$:
\begin{eqnarray}\label{Bnormcondition}
    \int D\tilde{Y}_{d} \tilde{P}_{d}[\tilde{Y}_{d}|X] \equiv \int D\vec{\chi}_{1}D\vec{\chi}_{2} \tilde{P}_{d}[\tilde{Y}_{d}|X] = 1.
\end{eqnarray}
The integrals in the eq. \eqref{Bnormcondition} are Gaussian ones and can be calculated using the formulas for the Gaussian integrals (see eqs. \eqref{Lformula1} and \eqref{Lformula2}) again. Splitting the normalization factor into the leading contribution ($\Lambda^{(0)}$) and the first correction in $\beta$ ($\Lambda^{(1)}$) we get for each order: 
\begin{eqnarray}\label{Bnormfactor0}
    \Lambda^{(0)} = \left(\frac{\pi Q L}{M\delta_{t}}\right)^{-M}
\end{eqnarray}
and 
\begin{eqnarray}\label{Bnormfactor1}
    &&\frac{\Lambda^{(1)}}{\Lambda^{(0)}} = \beta LTr\Big[\hat{b}_{1}^{L} + \hat{b}_{2}^{L} + \hat{b}_{4}^{L} \hat{D}_{1}^{L} + \hat{b}_{5}^{L} \hat{D}_{1}^{L} +\nonumber\\
    &&\hat{b}_{8}^{L} \hat{D}_{2}^{L} + \hat{b}_{9}^{L} \hat{D}_{2}^{L} \Big] + \beta LTr\Big[\hat{b}_{1}^{d} + \hat{b}_{2}^{d} + \hat{b}_{4}^{d} \hat{d}_{1} + \nonumber\\
    &&\hat{b}_{5}^{d} \hat{d}_{1} + \hat{b}_{8}^{d} \hat{d}_{2} + \hat{b}_{9}^{d} \hat{d}_{2}\Big].
\end{eqnarray}
After substituting contributions to the normalization factor from the eqs. \eqref{Bnormfactor0} and \eqref{Bnormfactor1} into the eq. \eqref{BPdtildeomega0} contribution of the high-frequency parts of the matrices $\hat{b}_{i}$ and $\hat{D}_{1,2}$ vanishes and we can write the conditional PDF in the following way: 
\begin{eqnarray}\label{BPdtildeomega1}
    &&\tilde{P}_{d}[\tilde{Y}_{d}(\omega)|X] = \tilde{\Lambda}_{d,\omega}^{(0)} \exp\left\{-\frac{M\delta_{t}}{QL}[\vec{\chi}_{1}^{\dag}\vec{\chi}_{1} + \vec{\chi}_{2}^{\dag}\vec{\chi}_{2}]\right\}\Big(1 + \frac{\tilde{\Lambda}_{d,\omega}^{(1)}}{\tilde{\Lambda}_{d,\omega}^{(0)}} -\nonumber\\
    &&\frac{\beta M \delta_{t}}{Q} \Big[\vec{\chi}_{1}^{\dag}\hat{b}_{1}^{d}\vec{\chi}_{1} +\vec{\chi}_{2}^{\dag}\hat{b}_{2}^{d}\vec{\chi}_{2} + \vec{\chi}_{1}^{\dag}\hat{b}_{3}^{d}\vec{\chi}_{2} + \vec{\chi}_{1}^{\dag}\hat{b}_{4}^{d}\hat{d}_{1}\vec{\chi}_{1} +\vec{\chi}_{2}^{\dag}\hat{b}_{5}^{d}\hat{d}_{1}\vec{\chi}_{2} + 
    \nonumber\\
    &&\vec{\chi}_{1}^{\dag}\hat{b}_{6}^{d}\hat{d}_{1}\vec{\chi}_{2} + \vec{\chi}_{2}^{\dag}\hat{b}_{7}^{d}\hat{d}_{1}\vec{\chi}_{1} +\vec{\chi}_{1}^{\dag}\hat{b}_{8}^{d}\hat{d}_{2}\vec{\chi}_{1} + 
    \vec{\chi}_{2}^{\dag}\hat{b}_{9}^{d}\hat{d}_{2}\vec{\chi}_{2} + \vec{\chi}_{1}^{\dag}\hat{b}_{10}^{d}\hat{d}_{2}\vec{\chi}_{2} +\nonumber\\
    &&\vec{\chi}_{2}^{\dag}\hat{b}_{11}^{d}\hat{d}_{2}\vec{\chi}_{1}\Big] \Big),
\end{eqnarray}
where we defined the normalization factor $\tilde{\Lambda}_{d,\omega} = \tilde{\Lambda}_{d,\omega}^{(0)}+\tilde{\Lambda}_{d,\omega}^{(1)}$ of the new PDF $\tilde{P}_{d}[\tilde{Y}_{d}(\omega)|X]$ as: 
\begin{eqnarray}
    &&\tilde{\Lambda}_{d,\omega}^{(0)} = \left(\frac{\pi Q L}{M \delta_{t}}\right)^{-M_{d}},\label{BLambdad0omega} \\ &&\frac{\tilde{\Lambda}_{d,\omega}^{(1)}}{\tilde{\Lambda}_{d,\omega}^{(0)}} = \beta LTr\Big[\hat{b}_{1}^{d} + \hat{b}_{2}^{d} + \hat{b}_{4}^{d} \hat{d}_{1}  + \hat{b}_{5}^{d} \hat{d}_{1} +
    \hat{b}_{8}^{d} \hat{d}_{2} + \hat{b}_{9}^{d} \hat{d}_{2} \Big]. \label{BLambdad1omega}
\end{eqnarray}
We wrote the argument of the PDF in the eq. \eqref{BPdtildeomega1} as $\tilde{Y}_{d}(\omega)$ and added the subscript $\omega$ to the normalization factor in order to specify that the PDF $\tilde{P}_{d}[\tilde{Y}_{d}(\omega)|X]$ is normalized by the condition \eqref{Bnormcondition}, where we integrated over the low-frequency Fourier harmonics of the functions $y_{1,2}(t)$. We can move to the time domain by introducing the DFT on a coarse grid with the number of points $M_{d}$
\begin{eqnarray}\label{Bcoarsefourier}
    \hat{y}_{i}(\omega_{n'}) = \frac{1}{M_{d}}\sum_{n=0}^{M_{d}-1} y_{i}(t_{n})e^{-\frac{2\pi i n n'}{M_{d}}}.
\end{eqnarray}
The discretization step $\tilde{\delta}_{t}$ of the coarse grid is defined by the following relation: 
\begin{eqnarray}
    T = M\delta_{t} = M_{d}\tilde{\delta}_{t}.
\end{eqnarray}
After substituting Fourier-harmonics from the eq$.$ \eqref{Bcoarsefourier} in the eq$.$ \eqref{BPdtildeomega1}, we get the conditional PDF in terms of the functions $y_{1,2}(t)$ defined on the coarse grid:
\begin{eqnarray}\label{BtildePDFshort}
    \tilde{P}_{d}[\tilde{Y}_{d}(\omega)|X] = \tilde{\Lambda}_{d,\omega}^{(0)}\left(1 + \frac{\tilde{\Lambda}_{d,\omega}^{(1)}}{\tilde{\Lambda}_{d,\omega}^{(0)}}\right)\exp\left\{-\frac{S_{coarse}^{(0)} + S_{coarse}^{(1)}}{Q}\right\},
\end{eqnarray}
where the functionals $S_{coarse}^{(0),(1)}$ are expressed through the values of the functions $b_{i}(t)$ and $y_{1,2}(t)$ on the coarse grid:
\begin{eqnarray}
    &&S_{coarse}^{(0)} = \frac{\tilde{\delta}_{t}}{L}\sum_{j=0}^{M_{d}-1} \left[y_{1}^{2}(t_{j}) + y_{2}^{2}(t_{j})\right], \label{BS_0coarse}\\
    &&S_{coarse}^{(1)} = \beta \tilde{\delta}_{t} \sum_{j=0}^{M_{d}-1} \{b_{1,j}y_{1,j}^{2} + b_{2,j}y_{2,j}^{2} + b_{3,j}y_{1,j}y_{2,j} + b_{4,j}y_{1,j}(\dot{y}_{1})_{j}  + \nonumber\\
    && b_{5,j}y_{2,j}(\dot{y}_{2})_{j} +b_{6,j}y_{1,j}(\dot{y}_{2})_{j} + b_{7,j}(\dot{y}_{1})_{j}y_{2,j} + b_{8,j}y_{1,j}(\ddot{y}_{1})_{j}+ \nonumber\\
    && b_{9,j}y_{2,j}(\ddot{y}_{2})_{j} + b_{10,j}y_{1,j}(\ddot{y}_{2})_{j} + b_{11,j}(\ddot{y}_{1})_{j}y_{2,j}\}.\label{BS_1coarse}
\end{eqnarray} 
However, the PDF from the eq$.$ \eqref{BtildePDFshort} is still normalized with respect to the Fourier harmonics of the functions $y_{1,2}(t)$. If we want our variables to be the values of the functions $y_{1,2}(t)$ on the coarse time grid, we also need to multiply the PDF $\tilde{P}_{d}[\tilde{Y}_{d}(\omega)|X]$ by the Jacobian determinant of the substitution \eqref{Bcoarsefourier}: 
\begin{eqnarray}
    \tilde{P}_{d}[\tilde{Y}_{d}(t)|X] = \left|\frac{\partial(\hat{y}_{1}(\omega), \hat{y}_{2}(\omega))}{\partial(y_{1}(t),y_{2}(t))}\right| \tilde{P}_{d}[\tilde{Y}(\omega)|X] = M_{d}^{-M_{d}} \tilde{P}_{d}[\tilde{Y}(\omega)|X]. 
\end{eqnarray}
Thus, we obtain the conditional PDF for the variables $y_{1,2}(t_{i})$: 
\begin{eqnarray}\label{BtildePDFshortt}
    \tilde{P}_{d}[\tilde{Y}_{d}(t)|X] = \tilde{\Lambda}_{d}^{(0)}\left(1 + \frac{\tilde{\Lambda}_{d}^{(1)}}{\tilde{\Lambda}_{d}^{(0)}}\right)\exp\left\{-\frac{S_{coarse}^{(0)} + S_{coarse}^{(1)}}{Q}\right\},
\end{eqnarray}
where the normalization factor reads: 
\begin{eqnarray}\label{BLambdadt}
    \tilde{\Lambda}_{d}^{(0)} = \left(\frac{\tilde{\delta}_{t}}{\pi Q L}\right)^{M_{d}}, \quad \frac{\tilde{\Lambda}_{d}^{(1)}}{\tilde{\Lambda}_{d}^{(0)}} = \frac{\tilde{\Lambda}_{d,\omega}^{(1)}}{\tilde{\Lambda}_{d,\omega}^{(0)}}.
\end{eqnarray}

The first correction to the normalization factor from the eq$.$ \eqref{BLambdad1omega} can be expressed through the functions $b_{i}(t)$. We recall that the matrix components $(\hat{b}_{i})_{nn'}$ were defined in the eq$.$ \eqref{Bbmatrix} as the Fourier transform of the function $b_{i}(t)$ evaluated at the frequency $\omega_{n} - \omega_{n'}$. Therefore we can immediately express the three terms which do not contain derivative operators ($\hat{b}_{1,2,3}^{d}$):
\begin{eqnarray}\label{BTrb}
    &&Tr\Big[\hat{b}_{i}^{d}\Big] = \sum_{n = n_{d}} (\hat{b}_{i})_{nn} = M_{d} \hat{b}_{i}(\omega = 0) = \frac{M_{d}}{M\delta_{t}} \sum_{n = 0}^{M-1} b_{i}(t_{n}) \delta_{t} \underset{\delta_{t}\rightarrow0}{\rightarrow}\nonumber\\
    &&M_{d} \int_{T} b_{i}(t) \frac{dt}{T},
\end{eqnarray}
where we denoted by $\sum_{n=n_{d}}$ the summation over such indices $n$ which correspond to the low frequencies $|\omega_{n}|\leq W_{d}/2$. Due to the fact that the trace is taken over the low-frequency indices, we can replace the elements of the matrices $\hat{d}_{1,2}$ with the elements of the derivative operator in the continuous case. Therefore, traces from the eq$.$ \eqref{BLambdad1omega} reduce to:
\begin{align}
    &Tr\Big[\hat{b}_{i}^{d}\hat{d}_{1}\Big] \rightarrow \frac{M_{d}}{W_{d}} \int_{T} b_{i}(t) dt \times \int_{-\frac{W_{d}}{2}}^{\frac{W_{d}}{2}} i\omega d\omega = 0,\label{Trbd1}\\
    &Tr\Big[\hat{b}_{i}^{d}\hat{d}_{2}\Big] \rightarrow - \frac{M_{d}}{W_{d}} \int_{T} b_{i}(t) dt \times \int_{-\frac{W_{d}}{2}}^{\frac{W_{d}}{2}} \omega^{2} d\omega = -M_{d} \frac{W_{d}^{2}}{12} \int_{T} b_{i}(t) \frac{dt}{T}.\label{Trbd2}
\end{align}
Explicit expressions for the functions $b_{i}(t)$ are rather cumbersome. Fortunately, because of the eq$.$ \eqref{Trbd1}, we only need to know expressions for $b_{1}(t)+b_{2}(t)$ and $b_{8}(t)+b_{9}(t)$ in order to get the normalization factor. Expressions for these quantities are relatively simple:
\begin{eqnarray}\label{Bsumsofbs}
    &&b_{1}(t) + b_{2}(t) = -\frac{\mu\dot{\mu}^{2}}{15(3+\mu^{2})^{3}(9+4\mu^{2})^{2}}(10206 + 21303\mu^{2}+
    15399\mu^{4}+\nonumber\\
    && 4644\mu^{6} +496\mu^{8})-\frac{4\mu\dot{\mu}\dot{\phi}}{3+\mu^{2}} -\frac{2(3+2\mu^{2})\ddot{\phi}}{3+\mu^{2}},\nonumber\\
    &&b_{8}(t)+b_{9}(t) = -\frac{4 \mu^{3}}{15(3+\mu^{2})}.
\end{eqnarray}
Finally, with the eq$.$ \eqref{Bsumsofbs} we get:
\begin{eqnarray}\label{Bfinalnormfactors}
    &&\frac{\tilde{\Lambda}_{d}^{(1)}}{\tilde{\Lambda}_{d}^{(0)}} = M_{d}\frac{\beta L W_{d}^{2}}{12} \int_{T} \frac{4 \mu^{3}}{15(3+\mu^{2})} \frac{dt}{T} - M_{d} \beta L \int  \Bigg[\frac{4\mu\dot{\mu}\dot{\phi}}{3+\mu^{2}} +\nonumber\\
    && \frac{2(3+2\mu^{2})\ddot{\phi}}{3+\mu^{2}} + \frac{\mu\dot{\mu}^{2}}{15(3+\mu^{2})^{3}(9+4\mu^{2})^{2}}(10206 + 21303\mu^{2} +\nonumber \\
    &&15399\mu^{4}+4644\mu^{6} +496\mu^{8})\Bigg]\frac{dt}{T}.
\end{eqnarray}
Averaging the eq$.$ \eqref{Bfinalnormfactors} with respect to the distribution of the input signal will give two contributions: the one proportional to the dimensionless parameter $\beta L W_{d}^{2}$ and the one proportional to $\beta L W_{X}^{2}$ (we get $W_{X}^{2}$ from averaging derivatives of the input signal).

\section{Appendix C}
In this section we explain in detail how we evaluated two averages from the section 5:
\begin{eqnarray}\label{Cintegrals}
    &&\int_{T} \frac{dt}{T}\langle f(\mu)\dot{\mu}^{2} \rangle_{P_{X}} = \int_{T} \frac{dt}{T} \int DX P_{X}[X] f(\mu)\dot{\mu}^{2},  \\ \nonumber
    &&\, \int_{T} \frac{dt}{T}\left\langle \frac{4 \mu^{3}}{15(3+\mu^{2})} \right\rangle_{P_{X}} = \int_{T} \frac{dt}{T} \int DX P_{X}[X] \frac{4 \mu^{3}}{15(3+\mu^{2})}.
\end{eqnarray}
In what follows we will work with real and imaginary parts of the input signal $X(t) = a(t)+i b(t)$, where $a$ and $b$ are the real functions of time. In the action $S[X]$ we have two independent oscillators in imaginary time with the same frequency: 
\begin{eqnarray}\label{Cactionab}
    &&S[X] = \int_{T} dt \left(\frac{m |\dot{X}|^{2}}{2}+\frac{m\Omega^{2}|X|^{2}}{2}\right) = \int_{T} dt \left(\frac{m \dot{a}^{2}}{2}+\frac{m\Omega^{2}a^{2}}{2}\right) \nonumber\\
    && + \int_{T} dt \left(\frac{m \dot{b}^{2}}{2}+\frac{m\Omega^{2}b^{2}}{2}\right).
\end{eqnarray}
Two-point correlation functions for the action \eqref{Cactionab} with the boundary conditions $X(T/2) = X(-T/2) = 0$ are well-known \cite{kleinert2009path}, and we will use it evaluating the averages from the eq$.$ \eqref{Cintegrals}: 
\begin{eqnarray}\label{correlatorscont}
    &&\langle a(t)a(s) \rangle_{P_{X}} = \langle b(t)b(s) \rangle_{P_{X}} =
    \begin{cases}
        \dfrac{\textrm{sh}\, \Omega (T/2+s)\, \textrm{sh}\, \Omega(T/2-t)}{m \Omega\, \textrm{sh}\, \Omega T}, & t>s\nonumber\\
        \\
        \dfrac{\textrm{sh}\, \Omega (T/2+t)\, \textrm{sh}\, \Omega(T/2-s)}{m \Omega\, \textrm{sh}\, \Omega T}, & t<s
    \end{cases},\\
    &&\langle a(t)b(s)\rangle_{P_{X}} = 0.
\end{eqnarray}

To evaluate path-integrals from the eq$.$ \eqref{Cintegrals} we introduce time discretization and the discrete version of the time derivative:
\begin{eqnarray}
    \dot{g}_{i}=\sum_{j=0}^{M}D_{ij}g_{j} = \dfrac{1}{\delta_{t}}\left(g_{i+1}-g_{i}\right), \, 0 \leq i \leq M-1,
\end{eqnarray}
where $g_{i} = g(t_{i}), \, \dot{g}_{i} = \dot{g}(t_{i})$  and $t_{i}$ is a point of the time grid with the discretization step $\delta_{t}$: $t_{i} = -T/2 + i\delta_{t}$. Due to the boundary condition $X(-T/2) = X(T/2) = 0$ we are considering functions $g(t)$ which vanish at the endpoints, i.e. $g_{M} = g_{0} = 0$.

We can write action from the eq$.$ \eqref{Cactionab} on the discretized time axis as: 
\begin{eqnarray}\label{oscaction}
    S[X] = \frac{1}{2}\sum_{j,k = 1}^{M-1} a_{j}Q_{jk}a_{k}+\frac{1}{2}\sum_{j,k=0}^{M-1} b_{j}Q_{jk}b_{k}, 
\end{eqnarray}
\begin{eqnarray}\label{matrixQ}
    Q_{jk} = m \delta_{t} \sum_{i=0}^{M-1}D_{ij}D_{ik} + m\Omega^{2}\delta_{t}\delta_{jk},
\end{eqnarray}
where $\delta_{t}$ is the discretization step. In terms of the discretized action the correlators from the eq$.$ \eqref{correlatorscont} are kernels of the integral operator $Q^{-1}$:
\begin{eqnarray}\label{disccorrelator}
    \langle a(t_{i})a(t_{j}) \rangle_{P_{X}} = \langle b(t_{i})b(t_{j}) \rangle_{P_{X}} = (Q^{-1})_{ij}.
\end{eqnarray}

Knowledge of the correlators from the eq$.$ \eqref{disccorrelator} and of the action from the eq$.$ \eqref{oscaction} is enough to evaluate average of the form $\langle f(\mu) \dot{\mu}^{2}\rangle_{P_{X}}$. First we recall how $\mu$ is related to the input signal: $\mu = \gamma L (a^{2}+b^{2})$, so $\dot{\mu}^{2} = 4(\gamma L)^{2}(a^{2}\dot{a}^{2}+b^{2}\dot{b}^{2}+2a b \dot{a}\dot{b})$, and we have three terms:
\begin{eqnarray}
    &&\langle f(\mu)\dot{\mu}^{2} \rangle_{P_{X}} = 4(\gamma L)^{2}\langle f(\mu) a^{2}\dot{a}^{2} \rangle_{P_{X}} + 4(\gamma L)^{2}\langle f(\mu) b^{2}\dot{b}^{2} \rangle_{P_{X}} + \nonumber\\
    &&8(\gamma L)^{2}\langle f(\mu) a b\dot{a}\dot{b} \rangle_{P_{X}}.
\end{eqnarray}
The first one reads: 
\begin{eqnarray}\label{firstaverage}
    \langle f(a_{i},b_{i})a_{i}^{2}\dot{a}_{i}^{2}\rangle_{P_{X}} = \sum_{j,k=1}^{M-1} D_{ij}D_{ik} \langle f(a_{i},b_{i})a_{i}^{2}a_{j}a_{k}\rangle_{P_{X}},
\end{eqnarray}
where we have moved to the discrete notation. In the eq$.$ \eqref{firstaverage} we do not imply summation over the repeated index $i$, and in what follows we always write sums explicitly. From the eq$.$ \eqref{firstaverage} we proceed by adding two additional integrals over delta-functions, which allows us to extract the function $f(a_{i},b_{i})$ from the average:
\begin{eqnarray}\label{firstaveragenoD}
    \langle f(a_{i},b_{i})a_{i}^{2}a_{j}a_{k}\rangle_{P_{X}} = \int dx dy f(x,y)x^{2}\langle \delta(x-a_{i})\delta(y-b_{i})a_{j}a_{k}\rangle_{P_{X}}.
\end{eqnarray}
Next we use integral representation of the delta-function:
\begin{eqnarray}\label{averagedeltaf}
    \langle \delta(x-a_{i})\delta(y-b_{i})a_{j}a_{k}\rangle_{P_{X}} = \int \frac{dk_{x}dk_{y}}{(2\pi)^{2}} \textrm{e}^{ik_{x}x+ik_{y}y}\langle \textrm{e}^{-ik_{x}a_{i}-ik_{y}b_{i}}a_{j}a_{k} \rangle_{P_{X}}.
\end{eqnarray}
Recalling the definition of the brackets $\langle ... \rangle_{P_{X}}$, we see that in case of the quadratic action from the eq$.$ \eqref{oscaction} the average from the eq$.$ \eqref{averagedeltaf} can be calculated explicitly. Namely, we change variables in the following integral: 
\begin{eqnarray}\label{oldintegral}
    \langle \textrm{e}^{-ik_{x}a_{i}-ik_{y}b_{i}}a_{j}a_{k} \rangle_{P_{X}} =\Lambda_{X} \int \left(\prod_{n} da_{n}db_{n}\right)  a_{j} a_{k} \textrm{e}^{-S[X]-ik_{x}a_{i}-ik_{y}b_{i}}
\end{eqnarray}
from $a_{j}$ and $b_{j}$ to $\alpha_{j} = a_{j}+ik_{x}(Q^{-1})_{ii}\delta_{ij}$ and $\beta_{j} = b_{j}+ik_{y}(Q^{-1})_{ii}\delta_{ij}$. After such substitution the integral from the eq$.$ \eqref{oldintegral} reads: 
\begin{eqnarray}
    \Lambda_{X}\int \left(\prod_{n} d\alpha_{n}d\beta_{n}\right) \left[\alpha_{j}\alpha_{k} - k_{x}^{2}(Q^{-1})_{ji}(Q^{-1})_{ki}\right]\textrm{e}^{-S[\alpha,\beta] - \frac{1}{2}(Q^{-1})_{ii}(k_{x}^{2}+k_{y}^{2})},
\end{eqnarray}
where $S[\alpha,\beta] = \frac{1}{2}\sum_{i,j=1}^{M-1}\alpha_{i}Q_{ij}\alpha_{j}+\frac{1}{2}\sum_{i,j=1}^{M-1}\beta_{i}Q_{ij}\beta_{j}$. Performing the Gaussian integration over $\alpha_{n}$ and $\beta_{n}$, we get: 
\begin{eqnarray}
    \langle \textrm{e}^{-ik_{x}a_{i}-ik_{y}b_{i}}a_{j}a_{k} \rangle_{P_{X}} = \left[(Q^{-1})_{jk}-k_{x}^{2}(Q^{-1})_{ji}(Q^{-1})_{ki}\right]\textrm{e}^{-\frac{1}{2}(Q^{-1})_{ii}(k_{x}^{2}+k_{y}^{2})},
\end{eqnarray}
where we also have used the expression for the normalization constant $\Lambda_{X} = \det
\left(\frac{Q}{2\pi}\right)$. Now we return to the eq$.$ \eqref{averagedeltaf} and integrate over $k_{x}$ and $k_{y}$ to get the following expression for the average $\langle \delta(x-a_{i})\delta(y-b_{i})a_{j}a_{k}\rangle_{P_{X}}$:
\begin{eqnarray}
    \frac{1}{2\pi(Q^{-1})_{ii}}\left[(Q^{-1})_{jk}-\frac{(Q^{-1})_{ji}(Q^{-1})_{ki}}{(Q^{-1})_{ii}}+x^{2}\frac{(Q^{-1})_{ji}(Q^{-1})_{ki}}{(Q^{-1})_{ii}^{2}}\right]\textrm{e}^{-\frac{x^{2}+y^{2}}{2(Q^{-1})_{ii}}}.
\end{eqnarray}
Finally, we arrive at the following result for one of the three terms which contribute to $\langle f(\mu)\dot{\mu}^{2} \rangle_{P_{X}}$:
\begin{eqnarray}\label{firsttermfmu}
    &&\langle f(a_{i},b_{i})a_{i}^{2}a_{j}a_{k}\rangle_{P_{X}} = \frac{1}{2\pi(Q^{-1})_{ii}}\int dxdy f(x,y)x^{2}\textrm{e}^{-\frac{x^{2}+y^{2}}{2(Q^{-1})_{ii}}} \nonumber\\
    &&\left[(Q^{-1})_{jk}-\frac{(Q^{-1})_{ji}(Q^{-1})_{ki}}{(Q^{-1})_{ii}}+x^{2}\frac{(Q^{-1})_{ji}(Q^{-1})_{ki}}{(Q^{-1})_{ii}^{2}}\right].
\end{eqnarray}

Evaluation of the other two terms ($\langle f(a,b)b^{2}\dot{b}^{2}\rangle_{P_{X}}$ and $2\langle f(a,b)a b \dot{a}\dot{b}\rangle_{P_{X}}$) can be done in the exactly same way by extracting the function $f(a,b)$ from the average sign by the virtue of introducing additional delta-functions. Moreover, we can get the answer for $\langle f(a_{i},b_{i})b_{i}^{2}b_{j}b_{k}\rangle_{P_{X}}$ by replacing $x$ with $y$ in the eq$.$ \eqref{firsttermfmu}: 
\begin{eqnarray}\label{secondtermfmu}
    &&\langle f(a_{i},b_{i})b_{i}^{2}b_{j}b_{k}\rangle_{P_{X}} = \frac{1}{2\pi(Q^{-1})_{ii}}\int dxdy f(x,y)y^{2}\textrm{e}^{-\frac{x^{2}+y^{2}}{2(Q^{-1})_{ii}}} \nonumber\\
    &&\left[(Q^{-1})_{jk}-\frac{(Q^{-1})_{ji}(Q^{-1})_{ki}}{(Q^{-1})_{ii}}+y^{2}\frac{(Q^{-1})_{ji}(Q^{-1})_{ki}}{(Q^{-1})_{ii}^{2}}\right].
\end{eqnarray}
Calculation of the last term $\langle f(a_{i},b_{i})a_{i}b_{i}a_{j}b_{k}\rangle_{P_{X}}$ gives us: 
\begin{eqnarray}\label{thirdtermfmu}
    &&\langle f(a_{i},b_{i})a_{i}b_{i}a_{j}b_{k}\rangle_{P_{X}} = \frac{1}{2\pi(Q^{-1})_{ii}}\int dxdy f(x,y)x^{2}y^{2}\textrm{e}^{-\frac{x^{2}+y^{2}}{2(Q^{-1})_{ii}}} \nonumber\\
    &&\frac{(Q^{-1})_{ji}(Q^{-1})_{ki}}{(Q^{-1})_{ii}^{2}}.
\end{eqnarray}
Now we combine all the three terms and restore the derivative operators $D_{ij}$, which we were omitting, to obtain $\langle f(\mu)\dot{\mu}^{2}\rangle_{P_{X}}$:
\begin{eqnarray}\label{faveragedisc}
    &&\langle f(\mu_{i}) \dot{\mu}_{i}^{2} \rangle_{P_{X}} = \sum_{j,k=1}^{M-1} \frac{4(\gamma L)^{2}}{2\pi (Q^{-1})_{ii}} \int dxdy f(x,y)(x^{2}+y^{2}) \textrm{e}^{-\frac{x^{2}+y^{2}}{2(Q^{-1})_{ii}}}D_{ij}D_{ik} \nonumber \\
    &&\left[(Q^{-1})_{jk}-\frac{(Q^{-1})_{ji}(Q^{-1})_{ki}}{(Q^{-1})_{ii}}+(x^{2}+y^{2})\frac{(Q^{-1})_{ji}(Q^{-1})_{ki}}{(Q^{-1})_{ii}^{2}}\right].       
\end{eqnarray}
We were able to express the average $\langle f(\mu)\dot{\mu}^{2}\rangle_{P_{X}}$ through the matrix $Q^{-1}$, which, as we recall from the eq$.$ \eqref{disccorrelator}, is essentially the two-point correlation function of our input signal. Next we should take the continuous limit of the expression \eqref{faveragedisc}, which amounts to the computation of derivatives of the correlation functions from the eq$.$ \eqref{correlatorscont}. However, if we try to evaluate the second derivative of the correlator at equal times (the term which contains $\sum_{j,k}D_{ij}D_{ik}(Q^{-1})_{jk}$), we will get a divergent result. It can be seen from the continuous expression for the second derivative of the correlator:
\begin{align}\label{seconddercorr}
    &\frac{d}{ds}\frac{d}{dt}\langle a(t) a(s)\rangle_{P_{X}} = \frac{\delta(t-s)}{m}-\nonumber\\
    &\frac{\Omega}{m\, \textrm{sh}\, \Omega T} \Big(\textrm{ch}\, \Omega (T/2+s)\, \textrm{ch}\, \Omega(T/2-t)+\textrm{ch}\, \Omega (T/2+t)\, \textrm{ch}\, \Omega(T/2-s)\Big).
\end{align}
To get the continuous limit of $D_{ij}D_{ik}(Q^{-1})_{jk}$ we are supposed to put $t=s$ in the eq$.$ \eqref{seconddercorr}. We see that due to the presence of the delta-function $\delta(t-s)$ we acquire a divergence trying to take the continuous limit.

We will deal with this divergence staying in discrete time by observing that the same singular term appears in the calculation of $W_{X}^{2}$. Therefore, we can express the singular term through $W_{X}^{2}$ and substitute it in $\langle f(\mu_{i})\dot{\mu}_{i}^{2}\rangle_{P_{X}}$, thus obtaining an expression which does not contain singular terms and allows us to take the continuous limit.
First of all, we should find an explicit expression for the matrix $Q^{-1}$. It can be done if we diagonalize the matrix $ (D^{T}D)_{jk} \equiv\sum_{i=0}^{M-1}D_{ij}D_{ik}$ (here $D^{T}$ stands for the transpose of the matrix $D$), which has the following form: 
\begin{align}\label{CDDmatrix}
    (D^{T}D)_{jk} = \dfrac{1}{\delta_{t}^{2}}
    \begin{pmatrix}
        2 & -1 & 0 & \cdots & 0 & 0 & 0 \\
        -1 & 2 & -1 & \cdots & 0 & 0 & 0 \\
        \vdots &&&&&&\vdots\\
        0 & 0 & 0 & \cdots & -1 & 2 & -1 \\
        0 & 0 & 0 & \cdots & 0 & -1 & 2 \\
    \end{pmatrix}.
\end{align}
The eigenvectors of the matrix from the eq$.$ \eqref{CDDmatrix} read:
\begin{align}\label{Cvvectors}
    v_{j}^{(\alpha)} = \sqrt{\dfrac{2}{M}}\,\textrm{sin}\,\dfrac{\pi j \alpha}{M}, \, 1 \leq \alpha \leq M-1,
\end{align}
where each vector corresponds to the eigenvalue $\nu_{\alpha}$:
\begin{align}\label{Cnualpha}
    \nu_{\alpha} = \dfrac{2}{\delta_{t}^{2}}\left(1 - \textrm{cos}\, \dfrac{\pi \alpha}{M}\right) = \dfrac{4}{\delta_{t}^{2}}\, \textrm{sin}^{2}\,\dfrac{\pi \alpha}{2 M}.
\end{align}
The normalization factor $\sqrt{2/M}$ in the eq$.$ \eqref{Cvvectors} ensures that the vectors $v_{j}^{(\alpha)}$ satisfy the following orthogonality and completeness relations: 
\begin{align}
    &\sum_{j=1}^{M-1} v_{j}^{(\alpha)}v_{j}^{(\beta)} = \delta_{\alpha \beta},\nonumber\\
    &\sum_{j=1}^{M-1} v_{j}^{(\alpha)}v_{k}^{(\alpha)} = \delta_{j k}.
\end{align}

We can expand the matrix $Q$ in terms of the vectors $v_{j}^{(\alpha)}$: 
\begin{align}
    Q_{jk} = m\delta_{t}(D^{T}D)_{jk}+m\Omega^{2}\delta_{t}\delta_{jk} = \sum_{\alpha=1}^{M-1} Q_{\alpha} v_{j}^{(\alpha)}v_{k}^{(\alpha)},
\end{align}
where the eigenvalues $Q_{\alpha}$ are:
\begin{align}
    Q_{\alpha} = m\delta_{t}\nu_{\alpha} + m \Omega^{2}\delta_{t}.
\end{align}
Therefore, the inverse of the matrix $Q$ is:
\begin{align}
    Q_{jk}^{-1} = \sum_{\alpha = 1}^{M-1} \dfrac{1}{Q_{\alpha}}v_{j}^{(\alpha)}v_{k}^{(\alpha)}.
\end{align}
Now we can compute the result of applying two derivative operators $D_{ij}$ to the matrix $Q^{-1}$:
\begin{align}\label{CDDonQ}
    \sum_{j,k=1}^{M-1} D_{ij}D_{ik}Q_{jk}^{-1} = \sum_{\alpha=1}^{M-1} \dfrac{1}{Q_{\alpha}} \left(\sum_{j=1}^{M-1}D_{ij}v_{j}^{(\alpha)}\right)^{2}.
\end{align}
Recalling the expressions for the components of the vectors $v_{j}^{(\alpha)}$ (see the eq$.$ \eqref{Cvvectors}), one can derive how $D_{ij}$ acts on these vectors:
\begin{align}\label{CDonvectors}
    \sum_{j=1}^{M-1} D_{ij}v_{j}^{(\alpha)} = \dfrac{2}{\delta_{t}}\sqrt{\dfrac{2}{M}}\,\textrm{sin}\,\dfrac{\pi \alpha}{2M}\,\textrm{cos}\,\dfrac{\pi \alpha(2i+1)}{2M},\, 0 \leq i \leq M-1,
\end{align}
and substitute the result of \eqref{CDonvectors} into the eq$.$ \eqref{CDDonQ}:
\begin{align}\label{Csingular}
     &\sum_{j,k=1}^{M-1} D_{ij}D_{ik}Q_{jk}^{-1} = \dfrac{8}{M\delta_{t}^{2}}\sum_{\alpha=1}^{M-1}\dfrac{1}{Q_{\alpha}}\,\textrm{sin}^{2}\dfrac{\pi \alpha}{2 M}\,\textrm{cos}^{2}\dfrac{\pi \alpha(2i+1)}{2M} =\\
     &\dfrac{2}{M}\sum_{\alpha=1}^{M-1}\dfrac{\nu_{\alpha}}{Q_{\alpha}}\,\textrm{cos}^{2}\dfrac{\pi \alpha(2i+1)}{2M} = \dfrac{2}{mT}\sum_{\alpha=1}^{M-1}\textrm{cos}^{2}\dfrac{\pi \alpha(2i+1)}{2M}-\nonumber\\
     &\dfrac{2\Omega^{2}}{mT}\sum_{\alpha=1}^{M-1}\dfrac{1}{\nu_{\alpha}+\Omega^{2}}\textrm{cos}^{2}\dfrac{\pi \alpha(2i+1)}{2M} = \dfrac{M-1}{mT}-\dfrac{2\Omega^{2}}{mT}\sum_{\alpha=1}^{M-1}\dfrac{1}{\nu_{\alpha}+\Omega^{2}}\textrm{cos}^{2}\dfrac{\pi \alpha(2i+1)}{2M}.
\end{align}

The squared bandwidth $W_{X}^{2}$ can be evaluated in the same way, which gives:
\begin{align}\label{Cbandwidthsing}
    &W_{X}^{2} = \frac{1}{P}\int_{T} \frac{dt}{T} \langle |\dot{X}|^{2} \rangle_{P_{X}} = \frac{1}{P} \sum_{i=0}^{M-1}\frac{\delta_{t}}{T} \langle \dot{a}^{2}_{i} + \dot{b}^{2}_{i}\rangle_{P_{X}} = \nonumber \\
    &\frac{2\delta_{t}}{PT}\sum_{i=0}^{M-1}\sum_{j,k=1}^{M-1}D_{ij}D_{ik}Q^{-1}_{jk} =  \dfrac{2\delta_{t}}{PT}\sum_{\alpha=1}^{M-1}\dfrac{\nu_{\alpha}}{Q_{\alpha}} = \dfrac{2(M-1)}{mPT} - \dfrac{2\Omega^{2}}{mPT}\sum_{\alpha=1}^{M-1}\dfrac{1}{\nu_{\alpha}+\Omega^{2}}.
\end{align}
We see that the expressions \eqref{Csingular} and \eqref{Cbandwidthsing} both contain the same singularity $M/T = 1/\delta_{t}  $. Expressing the singular term through the squared bandwidth and substituting it in the eq$.$ \eqref{Csingular}, we obtain:
\begin{align}\label{Csingularremoved}
    &\sum_{j,k=1}^{M-1} D_{ij}D_{ik}Q_{jk}^{-1} = \dfrac{1}{2}PW_{X}^{2} - \dfrac{\Omega^{2}}{mT}\sum_{\alpha=1}^{M-1}\dfrac{1}{\nu_{\alpha}+\Omega^{2}}\textrm{cos}\,\dfrac{\pi \alpha(2i+1)}{M}.
\end{align}

Now we can substitute the right-hand side of the eq. \eqref{Csingularremoved} in $\langle f(\mu_{i})\dot{\mu}^{2}_{i}\rangle_{P_{X}}$:
\begin{align}\label{3fren}
    &\langle f(\mu_{i}) \dot{\mu}_{i}^{2} \rangle_{P_{X}} =  \frac{4(\gamma L)^{2}}{2\pi (Q^{-1})_{ii}} \int dxdy f(x,y)(x^{2}+y^{2}) \textrm{e}^{-\frac{x^{2}+y^{2}}{2(Q^{-1})_{ii}}} \nonumber \\
    &\Bigg[\frac{1}{2}PW_{X}^{2}-\dfrac{\Omega^{2}}{mT}\sum_{\alpha=1}^{M-1}\dfrac{1}{\nu_{\alpha}+\Omega^{2}}\textrm{cos}\,\dfrac{\pi \alpha(2i+1)}{M}-\nonumber\\
    &\sum_{j,k=1}^{M-1}D_{ij}D_{ik}\frac{(Q^{-1})_{ji}(Q^{-1})_{ki}}{(Q^{-1})_{ii}}\left(1-\dfrac{x^{2}+y^{2}}{(Q^{-1})_{ii}}\right)\Bigg].
\end{align}
Before taking the continuous limit of the eq$.$ \eqref{3fren}, we should note that most of the terms in the eq$.$ \eqref{3fren} can be neglected, because we consider the amount of points $M \sim T W_{X}$ needed to represent the input signal to be large. Let us first consider the following sum:
\begin{align}\label{3firstsum}
    \dfrac{\Omega^{2}}{mT}\sum_{\alpha=1}^{M-1}\dfrac{1}{\nu_{\alpha}+\Omega^{2}}\textrm{cos}\,\dfrac{\pi \alpha(2i+1)}{M},
\end{align}
and compare it to $PW_{X}^{2}$. The contribution of the sum \eqref{3firstsum} can be estimated as follows:
\begin{align}
    &\dfrac{\Omega^{2}}{mT}\left|\sum_{\alpha=1}^{M-1}\dfrac{1}{\nu_{\alpha}+\Omega^{2}}\textrm{cos}\,\dfrac{\pi \alpha(2i+1)}{M}\right| \leq \dfrac{1}{mT}\sum_{\alpha=1}^{M-1} \dfrac{1}{1+(M/\xi)^{2}\textrm{sin}^{2}\,\dfrac{\pi \alpha}{2M}}\leq\nonumber\\ 
    & \dfrac{\xi^{2}}{m T M^{2}}\sum_{\alpha=1}^{M-1} \dfrac{1}{\textrm{sin}^{2}\,\dfrac{\pi \alpha}{2M}} \leq \dfrac{\xi^{2}}{m T}\sum_{\alpha=1}^{M-1}\dfrac{1}{\alpha^{2}} \leq \dfrac{2P \xi^{4}}{T^{2}}\dfrac{\pi^{2}}{6},
\end{align}
where we used the relations between $m$, $\Omega$ and $P$, $T$:
\begin{align}\label{3momegaPTrelations}
    \dfrac{1}{P\xi} = m\Omega, \, 2\xi = \Omega T,
\end{align}
and also the following inequality:
\begin{align}
    \textrm{sin}\,x \geq \dfrac{2}{\pi}x, \qquad |x| \leq \pi/2.
\end{align}
Now one can see that the sum \eqref{3firstsum} is negligible, because $W_{X} \gg 1/T$. The second term which should be considered reads:
\begin{align}\label{3secondsum}
    &\sum_{j,k=1}^{M-1}D_{ij}D_{ik}\frac{(Q^{-1})_{ji}(Q^{-1})_{ki}}{(Q^{-1})_{ii}} = \dfrac{4 P M^{2}}{T^{2}}\Bigg(\sum_{\alpha=1}^{M-1}\dfrac{1}{1+(M/\xi)^{2}\textrm{sin}^{2}\,\dfrac{\pi \alpha}{2M}}\textrm{sin}\,\dfrac{\pi \alpha}{2M}\nonumber\\
    &\textrm{cos}\,\dfrac{\pi \alpha(2i+1)}{2M}\textrm{sin}\dfrac{\pi i \alpha}{M}\Bigg)^{2}\Bigg /\Bigg(\sum_{\alpha=1}^{M-1}\dfrac{1}{1+(M/\xi)^{2}\textrm{sin}^{2}\,\dfrac{\pi \alpha}{2M}}\textrm{sin}^{2}\,\dfrac{\pi i \alpha}{M}\Bigg) \leq\nonumber\\
    & \dfrac{A P M}{T^{2}} \sim \dfrac{A P W_{X}}{T},
\end{align}
where $A$ is some real positive constant. We see again that the contribution from the eq$.$ \eqref{3secondsum} can be neglected due to the presence of the factor $1/T \ll W_{X}$.

Finally, neglecting the small terms and changing variables in the integral to $x=\rho\, \textrm{cos}\phi$ and $y = \rho\, \textrm{sin} \phi$, we write the continuous limit of the eq$.$ \eqref{3fren}: 
\begin{align}\label{faveragecont}
    &\langle f(\mu(t))\dot{\mu}(t)^{2} \rangle_{P_{X}} \approx \frac{2PW_{X}^{2}(\gamma L)^{2}m\Omega\,\textrm{sh}\, \Omega T}{\textrm{sh}\, \Omega (T/2+t) \, \textrm{sh}\, \Omega (T/2-t)} \int_{0}^{+\infty}d\rho \rho^{3}f(\mu)\\\nonumber
    &\exp{-\frac{m\Omega \rho^{2}}{2}\frac{\textrm{sh}\, \Omega T}{\textrm{sh}\, \Omega (T/2+t) \, \textrm{sh}\, \Omega (T/2-t)}}. 
\end{align}

Integration in the eq$.$ \eqref{faveragecont} over the time interval $[-T/2,T/2]$ will give us a contribution to the mutual information. The integral over time can be calculated by first shifting the integration interval: $t = t' - T/2$ and then changing the variable from $t'$ to $x = \left(\textrm{cth}\,\Omega(T-t')-\textrm{cth}\,\Omega T\right)^{-1}$. Then we have the following integral: 
\begin{eqnarray}
    &&\int_{0}^{T} \frac{dt'}{T}\langle f(\mu)\dot{\mu}^{2} \rangle_{P_{X}} \approx \frac{2mPW_{X}^{2}(\gamma L)^{2}}{T} \int_{0}^{+\infty} d\rho f(\mu)\rho^{3}\int_{0}^{+\infty}\frac{dx}{x}\\ \nonumber
    &&\textrm{exp}\left\{-\frac{m\Omega\rho^{2}}{2}\left(\frac{1}{x}+\frac{x}{\textrm{sh}^{2}\,\Omega T}\right)-\textrm{cth}\,\Omega T\,m\Omega\rho^{2}\right\},
\end{eqnarray}
which can be calculated by using the integral representation of the modified Bessel function $K_{\nu}(x)$:
\begin{eqnarray}
    \int_{0}^{+\infty}dx x^{\nu-1}\textrm{exp}\left(-\frac{p}{x}-q x\right) = 2\left(\frac{p}{q}\right)^{\nu/2} K_{\nu}(2\sqrt{pq}), \, p>0, \, q>0.
\end{eqnarray}
Eventually, we get the following result:
\begin{align}
    &\int_{-T/2}^{T/2} \frac{dt}{T}\langle f(\mu)\dot{\mu}^{2} \rangle_{P_{X}} \approx \frac{4(\gamma L)^{2}m P W_{X}^{2}}{T}\int_{0}^{+\infty} d\rho f(\mu)\nonumber\\
    &\rho^{3}\textrm{e}^{-m\Omega\textrm{cth}\,\Omega T\,\rho^{2}}K_{0}\left(\frac{m\Omega\rho^{2}}{\textrm{sh}\,\Omega T}\right).
\end{align}

Expressing $m$ and $\Omega$ through the parameters of the input signal $P$ and $W_{X}^{2}$ according to the relations \eqref{3momegaPTrelations} we arrive at the formula:
\begin{align}
    &\int_{-T/2}^{T/2} \frac{dt}{T}\langle f(\mu)\dot{\mu}^{2} \rangle_{P_{X}} \approx (\gamma L P )^{2} W_{X}^{2}\textrm{sh}^{2}\, 2\xi\int_{0}^{+\infty}dy y f\left(\gamma L P\xi\, \textrm{sh}\,2\xi \,y\right)\\ \nonumber
    &K_{0}\left(y\right)\textrm{e}^{-y\, \textrm{ch}\,2\xi}.
\end{align}

The second integral which we have to evaluate can be treated in the exactly same way as the first integral. The result reads: 
\begin{align}
    &\int_{-T/2}^{T/2} \frac{dt}{T}\left\langle \frac{4 \mu^{3}}{15(3+\mu^{2})} \right\rangle_{P_{X}} = \frac{2}{15}\xi^{2}(\gamma L P)^{3}\textrm{sh}^{4}\,2\xi\nonumber\\
    &\int_{0}^{+\infty}dy \frac{y^{3}K_{0}\left(y\right)}{3+(\gamma L P\xi)^{2}\textrm{sh}^{2}\,2\xi \,y^{2}}\textrm{e}^{-y\,\textrm{ch}\,2\xi}
\end{align}

\bibliographystyle{IEEEtran}
\bibliography{biblio.bib}
\end{document}